\documentclass{sig-alternate-05-2015}

\usepackage{graphicx}
\usepackage{cite}
\usepackage{caption}
\usepackage{subcaption}
\usepackage{amssymb}
\usepackage{amsmath}
\usepackage{amsfonts}
\usepackage[english]{babel}
\usepackage[bottom]{footmisc}
\usepackage{wrapfig}
\usepackage{commath}
\usepackage{algorithm}
\usepackage{wrapfig}
\usepackage[noend]{algpseudocode}
\usepackage{algorithm}
\usepackage{algpseudocode}
\usepackage[noend]{algpseudocode}
\usepackage{algorithm}
\usepackage{algpseudocode}
\usepackage{xspace}
\usepackage{amsmath,amsfonts,amssymb,amsthm}
\usepackage{colortbl}
\usepackage[T1]{fontenc}
\usepackage{bbm}
\usepackage[colorinlistoftodos,prependcaption,textsize=tiny]{todonotes}        

\renewcommand{\phi}{\varphi}

\newcommand{\ltl}{\mbox{\text{cs}\textsc{ltl}}\xspace}
\newcommand{\sstl}{\mbox{\textsc{bltl}}\xspace}
\newcommand{\kkt}{\mbox{\textsc{kkt}}\xspace}

\newcommand{\reals}{\mathbb{R}}
\newcommand{\naturals}{\mathbb{N}}
\newcommand{\integers}{\mathbb{Z}}

\newcommand{\integersgz}{\integers_{\geq 0}}
\newcommand{\N}{\mathcal{N}}

\newcommand{\U}{A}
\renewcommand{\S}{\mathcal{S}}

\renewcommand{\L}{{J}}
\newcommand{\AP}{\Upsilon}
\newcommand{\ap}{p}
\newcommand{\transfMatrix}{\mathcal{T}}

\newcommand{\pX}{\mathbf{x}}

\newcommand{\trace}{\xi}

\newcommand{\cov}{{Cov}}

\newcommand{\safeSet}{X}

\newcommand{\discreteKernel}{T}

\newcommand{\mdp}{\mbox{\textsc{mdp}}\xspace}
\newcommand{\imdp}{\mbox{\textsc{imdp}}\xspace}
\newcommand{\M}{\mathcal{M}}
\newcommand{\I}{\mathcal{I}}

\newcommand{\probDist}{\mathcal{D}}
\newcommand{\FeasibleDist}[2]{\Gamma_{#1}^{#2}}
\newcommand{\feasibleDist}[2]{\gamma_{#1}^{#2}}

\newcommand{\Amdp}{A}
\newcommand{\Qmdp}{Q}
\newcommand{\Pmdp}{P}
\newcommand{\amdp}{a}
\newcommand{\qmdp}{q}
\newcommand{\qmdpprime}{\qmdp^\prime}
\newcommand{\Qimdp}{Q}
\newcommand{\Aimdp}{A}
\newcommand{\qimdp}{q}
\newcommand{\qimdpprime}{\qimdp^\prime}
\newcommand{\aimdp}{a}
\newcommand{\Fimdp}{\Qimdp_{\bar{\phi}\mathrm{ac}}}

\newcommand{\shs}{\mbox{\textsc{shs}}\xspace}

\renewcommand{\H}{\mathcal{H}}

\newcommand{\HybridSystem}{\mathcal{H}}

\newcommand{\HybridSpace}{S}
\newcommand{\HybridState}{s}
\newcommand{\pH}{\mathbf{\HybridState}}
\newcommand{\pA}{\mathbf{a}}
\newcommand{\pathH}[1]{\omega_{\HybridSystem}^{#1}}
\newcommand{\PathH}{\mathit{Paths}_{\HybridSystem}}
\newcommand{\PathHK}{\mathit{Paths}^{\kappa}_{\HybridSystem}}
\newcommand{\PathHfin}{\mathit{Paths}^{\mathrm{fin}}_{\HybridSystem}}

\newcommand{\Pup}{\hat{P}}
\newcommand{\Plow}{\check{P}}

\newcommand{\pathmdp}{\omega}
\newcommand{\pathimdp}{\omega}
\newcommand{\pathmdpfin}{\pathmdp^{\mathrm{fin}}}
\newcommand{\pathimdpfin}{\pathimdp^{\mathrm{fin}}}
\newcommand{\Pathmdp}{\mathit{Paths}}
\newcommand{\Pathmdpfin}{\mathit{Paths}^{\mathrm{fin}}}
\newcommand{\Pathimdp}{\mathit{Paths}}
\newcommand{\Pathimdpfin}{\mathit{Paths}^{\mathrm{fin}}}
\newcommand{\last}{\mathit{last}}

\newcommand{\str}{\sigma}
\newcommand{\Str}{\Sigma}
\newcommand{\adv}{\pi}

\newcommand{\strH}{\str_{\HybridSystem}}

\newcommand{\dfa}{\textsc{dfa}\xspace}
\newcommand{\Adfa}{\mathcal{A}}
\newcommand{\Zdfa}{Z}
\newcommand{\zdfa}{z}

\newcommand{\Trandfa}{\tau}
\newcommand{\Fdfa}{\Zdfa_\mathrm{ac}}

\newcommand{\prob}{\mathit{Prob}}
\newcommand{\safe}{\mathrm{safe}}
\newcommand{\unsafe}{\mathrm{u}}

\newcommand{\Post}{\mathit{Post}}
\newcommand{\conv}{\mathit{conv}}
\newcommand{\erf}{\mathrm{erf}}

\newcommand{\Id}{\mathbf{I}}

\newcommand{\p}{p}
\newcommand{\plow}{\check{\p}}
\newcommand{\pup}{\hat{\p}}

\newcommand{\err}{\varepsilon}

\newcommand{\dx}{\Delta x}

\newtheorem{theorem}{Theorem}
\newtheorem{corollary}{Corollary}

\newtheorem{mydef}{Definition}
\newtheorem{myprob}{Problem}
\newtheorem{lemma}{Lemma}
\newtheorem{proposition}{Proposition}

\newtheorem{remark}{Remark}

\graphicspath{{figures/}}

\begin{document}
%
\title{Efficiency through Uncertainty: Scalable Formal Synthesis for Stochastic Hybrid Systems}
%
%
%

\author{Nathalie~Cauchi,
        Luca~Laurenti,
        Morteza~Lahijanian,\\
        Alessandro~Abate,
        Marta~Kwiatkowska,
        and~Luca~Cardelli%
\thanks{L. Laurenti, 
        M. Lahijanian, 
        A. Abate,  
        L. Cardelli, 
        and
        M. Kwiatkowska 
        are with the Dept. of Computer Science at University of Oxford, U.K. (email: \{\textit{firstname}.\textit{lastname}\}@cs.ox.ac.uk). L. Cardelli is also with Microsoft Research, Cambridge, UK.
        This work was supported in part by EPSRC Mobile Autonomy Program Grant EP/M019918/1, Royal Society grant RP120138, Malta's ENDEAVOUR scholarship scheme and  the Turing Institute, London, UK.}
}
\maketitle

\begin{abstract}

This work targets the development of an efficient abstraction method for formal analysis and control synthesis of discrete-time \textit{stochastic hybrid systems} (\shs) with linear dynamics. 
The focus is on temporal logic specifications, both over finite and infinite time horizons.  
The framework constructs a finite abstraction as a class of uncertain Markov models known as \emph{interval Markov decision process} (\imdp). 
Then, a strategy that maximizes the satisfaction probability of the given specification is synthesized over the \imdp and mapped to the underlying \shs.  
In contrast to existing formal approaches, which are by and large limited to finite-time properties and rely on conservative over-approximations, we show that the exact abstraction error can be computed as a solution of convex optimization problems  
and can be embedded into the \imdp abstraction. This is later used in the synthesis step over both finite- and infinite-horizon specifications, mitigating the known state-space explosion problem. Our experimental validation of the new approach compared to existing abstraction-based approaches shows: (i) significant (orders of magnitude) reduction of the abstraction error; (ii) marked speed-ups; and (iii) boosted scalability, allowing in particular to verify models with more than 10 continuous variables.  
\end{abstract}

\section{Introduction}
  \label{sec:introduction}
%
%
%
%

\textit{Stochastic hybrid systems} ({\shs}) are general and expressive models for the quantitative description of complex dynamical and control systems, such as cyber-physical systems. 
\shs have been used for modeling and analysis in diverse domains, 
ranging from avionics \cite{BL06} 
to chemical reaction networks \cite{cardelli2016stochastic} and manufacturing systems \cite{CL06}.
Many of these applications are \emph{safety-critical}; as a consequence,  
a theoretical framework providing formal guarantees for analysis and control of \shs is of major importance. 

Formal verification and synthesis for stochastic processes and \shs have been the focus of many recent studies  \cite{esmaeil2013adaptive,laurenti2017reachability,lahijanian2015formal,vinod2017forward,zamani2014symbolic}.   
These methods can provide formal guarantees on the probabilistic satisfaction of quantitative specifications, 
such as those expressed in \textit{linear temporal logic} (\textsc{ltl}). 
An approach to formal verification, 
which is particularly relevant for discrete-time models, 
hinges on the abstraction of continu\allowbreak{ous}-space stochastic models into discrete-space Markov process  \cite{Lahijanian:CDC:2012,esmaeil2013adaptive,lahijanian2015formal}. 
This leads to discrepancies between the abstract and original models, which can be captured through error guarantees. 
The main issue with this approach is its lack of scalability to complex models, which is related to the known state-space explosion problem.  This issue is aggravated by the conservative nature of the error bounds; thus,  to guarantee a given verification error, a very fine abstraction is generally required, leading to state-space explosion.  

This paper introduces a theoretical and computational synthesis framework for discrete-time \shs that is both formal and scalable. 
We zoom in on \shs that take the shape of switching diffusions \cite{yin2010hybrid}, 
which are linear in the continuous dynamics and where the control action resides in a mode switch. 
We focus on two fragments of \textsc{ltl} to encode properties for the \shs, namely \textit{co-safe} \textsc{ltl} (\ltl) \cite{kupferman:FMSD:2001}, which allows the expression of unbounded and complex reachability properties, and \textit{bounded} \textsc{ltl} (\sstl) \cite{jha2009bayesian}, which enables the expression of bounded-time and safety properties. 
The framework consists of two stages (abstraction and control synthesis) and puts forward key novel contributions.   
In the first step, 
(i) we introduce a novel space discretization technique that is dynamics-dependent, 
and (ii) we derive an analytical form for tight (exact) error bounds between the abstraction and the original model, 
(iii) which is reduced to the solution of a set of convex optimization problems leading to fast computations. 
The error is formally embedded as uncertain transition probabilities in the abstract model. 
In the second stage, 
(iv) a strategy (control policy) is computed by considering only feasible transition probability distributions over the abstract model, preventing the explosion of the error term.  
Finally, this strategy is soundly refined to a switching strategy for the underlying \shs with guarantees on the computed probability bounds. 
We provide (v) an illustration of the efficacy of the framework via three case studies, including a comparison with the state of the art. 

In conclusion, this work provides a new computational abstraction framework for discrete-time \shs that is both formal and markedly more scalable than state-of-the-art techniques and tools.  

\section{Problem Formulation}
  \label{sec:problemform}

We consider a \shs and a property of interest given as a temporal logic statement.  We are interested in computing a switching strategy for this model that optimizes the probability of achieving the property.  Below, we formally introduce the model, property, and problem.

\subsection{Stochastic Hybrid Systems}
\label{sec:switcheddiff}

We consider a class of discrete-time \shs with linear continuous dynamics and no resets of the continuous components.
\begin{mydef}[\shs] \label{def:shs}
    A (discrete-time) linear stochastic hybrid system $\HybridSystem$ is a tuple $\HybridSystem=(\U,F,G,\AP,L)$, where
    \begin{itemize}
    	\setlength\itemsep{0.5mm}
        \item $\U = \{a_1,\ldots,a_{|\U|}\}$ is a finite set of discrete modes, each of which containing a continuous domain $
        \reals^m$, defining the hybrid state space $\HybridSpace = \U \times \reals^{m}$,
        \item $F=\{F(a)\in \reals^{m \times m} \mid a \in \U \}$ is a collection of drift terms,
        \item $G=\{G(a)\in \reals^{m \times r} \mid a \in \U \}$ is a collection of diffusion terms,
        \item $\AP = \set{\ap_1,\ldots,\ap_n}$ is a set of atomic propositions,
        \item $L : S \to 2^\AP$ is a labeling function that assigns to each hybrid state possibly several elements of $\AP$. 
    \end{itemize}
\end{mydef}

\noindent

A pair $s = (a,x) \in \HybridSpace$, where $a \in \U$ and $x\in 
\reals^m$, denotes a hybrid state of $\HybridSystem$, and the evolution of $\HybridSystem$ for $k \in \integersgz$ is a stochastic process $\pH(k)=(\pA(k),\pX(k))$ with values in $\HybridSpace$.
The term $\pX$ represents the evolution of the continuous component of $\HybridSystem$ according to the stochastic difference equation
\begin{eqnarray}
\label{eq:switched-dynamics}
    &\pX(k+1)=F(a) \pX(k) + G(a) w,& \\
    &a \in \U, \quad w \sim \N(0,\cov_w),& \nonumber
\end{eqnarray}
where $w$ is a Gaussian noise with zero mean and covariance matrix $\cov_w \in \reals^{r \times r}$.  
The signal $\pA$ describes the evolution of the discrete modes over time.  

For $\kappa \in \integersgz \cup \{\infty\}$,  we call $\PathHK: \{0,1,\ldots,\kappa \} \to S$ the set of sample paths of $\pH$ of length $\kappa$.  The set of all sample paths with finite and infinite lengths are denoted by $\PathHfin$ and $\PathH$.  We denote by $\pathH{}$, $\pathH{k}$, and $\pathH{}(i)$ a sample path, a sample path of length $k$, and the $(i+1)$-th state on the path $\pathH{}$ of $\HybridSystem$, respectively.

\begin{mydef}[Switching strategy]
    \label{def:switching-strategy}
    A \emph{switching strategy} for $\HybridSystem$ is a function $\strH: \PathHfin \to \U$ that assigns a discrete mode $a \in \U$ to a finite path $\pathH{}$ of the process $\pH$.  The set of all switching strategies is denoted by $\Str_\H$.
\end{mydef}
\noindent
Given a switching strategy $\strH$, the evolution of $\pH(k)$ for $k < \kappa$,  is defined on the probability space $(\HybridSpace^{\kappa+1},\mathcal{B}(\HybridSpace^{\kappa +1}),P)$,
where $\mathcal{B}(\HybridSpace^{\kappa +1})$ is the product sigma-algebra on the product space $\HybridSpace^{\kappa +1}$, and $P$ is a probability measure.
We call $\discreteKernel$ the \emph{transition kernel} such that for any measurable set $B \subseteq \reals^m$, $x\in \reals^m$, and $a \in \U$, 
\begin{align}
\label{Eq:TransitionKernel}
	\discreteKernel(B \mid x,a)
				   			   =& \int_B  \N (t \mid F(a) x, G(a)^T \cov_w G(a)  ) \,dt.
\end{align} 
Then, it holds that $P$ is uniquely defined by $\discreteKernel$, and for $k< \infty$, 
\[\discreteKernel(B \mid x_k,a_k)=P(\pX(k+1) \in B \mid \pX(k)=x_k, \pA(k)=a_k).\] 
We note that, for $\kappa=\infty$, $P$ is still uniquely defined by $\discreteKernel$ by the \emph{Ionescu-Tulcea extension theorem} \cite{abate2014effect}.

We are interested in the properties of $\HybridSystem$ in set $(\U \times X) \subset S$, where $X \subset \reals^m$  is a continuous compact set.  Specifically, we analyze the behavior of $\HybridSystem$ with respect to a set of closed regions of interest $R = \set{r_1,\ldots,r_n}$, where $r_i \subseteq X$.  To this end, we associate to each region $r_i$ the atomic proposition (label) $\ap_i$, i.e., 
$\ap_i \in L(s = (a,x)) \Leftrightarrow x \in r_i$. 
Further, we define the (observation) \textit{trace} of path $\pathH{k} = s_0 s_1 \ldots s_k$ to be
\begin{equation*}
	\trace = \trace_0 \trace_1 \ldots \trace_k,
\end{equation*}
where $\trace_i = L(s_i) \in 2^{\AP}$ for all $i \leq k$.  For a path $\pathH{} \in \PathH$ with infinite length, we obtain an infinite-length trace.



\subsection{Temporal Logic Specifications}
\label{sec:temporal-logic}
We employ \textit{co-safe linear temporal logic} (\ltl) \cite{kupferman:FMSD:2001} and \textit{bounded linear temporal logic} (\sstl) \cite{jha2009bayesian} to write the properties of $\HybridSystem$.  We use \ltl to encode complex reachability properties with no time bounds, and \sstl to specify bounded-time properties. 

\begin{mydef}[\ltl syntax] 
	\label{def:sytax_scltl}
	A \ltl formula $\phi$ over a set of atomic proposition $\AP$ is inductively defined as follows:
	\begin{eqnarray*}
		\phi := \ap \,|\, \neg \ap \,|\, \phi \vee \phi \,|\, \phi \wedge \phi \,|\, \mathcal{X} \phi \,|\, \phi \,\mathcal{U} \phi \,|\, \mathcal{F} \, \phi,  
	\end{eqnarray*}
	where $\ap \in  \AP$, $\neg$ (negation), $\vee$ (disjunction), and $\wedge$ (conjunction) are Boolean operators, and $\mathcal{X}$ (``next"), $\mathcal{U}$ (``until"), and $\mathcal{F}$ (``eventually") are temporal operators.  
\end{mydef}




\begin{mydef}[\sstl syntax]
	\label{sstl-syntax}
	A \sstl formula $\phi$ over a set of atomic proposition $\AP$ is inductively defined as following:
	\begin{equation*}
		\phi := p \:|\: \neg \phi \:|\: \phi \, \vee \phi \:|\: \mathcal{X} \phi \:|\: \phi \, \mathcal{U}^{\leq k} \phi \:|\: \mathcal{F}^{\leq k} \phi \:|\: \mathcal{G}^{\leq k} \phi,
	\end{equation*}
	where $\ap \in \AP$ is an atomic proposition, $\neg$ (negation) and $\vee$ (disjunction) are Boolean operators, $\mathcal{X}$ (``next''), $\mathcal{U}^{\leq k}$ (``bounded until"), $\mathcal{F}^{\leq k}$ (``bounded eventually"), and $\mathcal{G}^{\leq k}$ (``bounded always") 
	are temporal operators.
\end{mydef}



\begin{mydef}[Semantics] \label{def:sematics}
	The semantics of \ltl and \sstl path formulas are defined over infinite traces over $2^\AP$.  Let $\trace = \{\trace_i\}_{i=0}^{\infty}$ with $\trace_i \in 2^\AP$ be an infinite trace and $\trace^i=\trace_i \trace_{i+1} \ldots$ be the $i$-th suffix.  Notation $\trace \models \phi$ indicates that $\trace$ satisfies formula $\phi$ and is recursively defined as following: 
	\begin{itemize}
		\setlength\itemsep{0.25mm}
		\small{
		\item $\trace \models \ap$ \:\text{if}\: $\ap \in \trace_0$; 
		\item $\trace \models \neg \phi$ \:if\: $\trace \not\models \phi$;
		\item $\trace \models \phi_1 \vee \phi_2$ \:if\: $\trace \models \phi_1$ or $\trace \models \phi_2$;
		\item $\trace \models \phi_1 \wedge \phi_2$ \:if\: $\trace \models \phi_1$ and $\trace \models \phi_2$;
		\item $\trace \models \mathcal{X} \phi$ \:if\: $\trace^1 \models \phi$;
		\item $\trace \models \phi_1 \mathcal{U} \phi_2$ \:if\: $\exists k \geq 0$, $\trace^k \models \phi_2$, and $\forall i \in [0,k)$, $\trace^i \models \phi_1$;
		\item $\trace \models \mathcal{F} \phi$ \:if\: $\exists k \geq 0$, $\trace^k \models \phi$;
		\item $\trace \models \phi_1 \mathcal{U}^{\leq k} \phi_2$ \:if\: $\exists j \leq k$, $\trace^j \models \phi_2$, and $\forall i [0,j)$, $\trace^i \models \phi_1$;
		\item $\trace \models \mathcal{F}^{\leq k} \phi$ \:if\: $\exists j \leq k$, $\trace^j \models \phi$;
		\item $\trace \models \mathcal{G}^{\leq k} \phi$ \:if\: $\forall j \leq k$, $\trace^j \models \phi$.
		}
	\end{itemize}
\end{mydef}
 
A trace $\trace$ satisfies a \ltl or \sstl formula $\phi$ iff there exists a ``good'' finite prefix $\underline{\trace}$ of $\trace$ such that the concatenation $\underline{\trace}\overline{\trace}$ satisfies $\phi$ for every suffix $\overline{\trace}$ \cite{kupferman:FMSD:2001,jha2009bayesian}.  
Therefore, even though the semantics of \ltl and \sstl are defined over infinite traces, 
we can restrict the analysis to the set of their good prefixes, which consists of finite traces.

\subsection{Problem Statement}
\label{sec:problem}

We say that a finite path $\pathH{}$ of $\HybridSystem$, initialized at state $s_0 \in S$, satisfies a formula $\phi$ if the path remains in the compact set $X$ and its corresponding finite trace $\trace \models \phi$.  
Under a switching strategy $\strH$, the probability that the \shs satisfies $\phi$ is given by:
\begin{multline}
	\label{eq:prob-phi}
	P(\phi \mid s_0, X, \strH) = P \big( \pathH{} \in \, \PathH^{\text{fin},\strH} \mid \pathH{}(0) = s_0,  \\
	\pathH{}(k) \in (\U \times X) \; \forall k \in [0, |\pathH{}|\,], \, \trace \models \phi \big), 
\end{multline}
where $\PathH^{\text{fin},\strH}$ denotes the set of all finite paths under strategy $\strH$, and $\trace$ is the observation trace of $\pathH{}$.
In this work, we are interested in synthesizing a switching strategy that maximizes the probability of satisfying property $\phi$.  

\begin{myprob}[Strategy synthesis]
\label{Problem 1}
	Given the $\shs$ $\HybridSystem$ in Def. \ref{def:shs}, a continuous compact set $X$
	, and a property expressed as a \ltl or \sstl formula $\phi$, find a switching strategy $\strH^*$ that maximizes the probability of satisfying $\phi$
	\begin{equation*}
		\strH^* = \arg \max_{\strH \in \Str_\H} P(\phi \mid s_0, X, \strH)
	\end{equation*}
	for all initial states $s_0 \in \U \times X$. 
\end{myprob}


\subsection{Overview of Proposed Approach}
\label{sec:approach}

We solve Problem \ref{Problem 1} with a discrete abstraction that is both formal and computationally tractable.  We construct a finite model in the form of an uncertain Markov process that captures all possible behaviors of the \shs $\HybridSystem$.  This construction involves a discretization of the continuous set $X$ and hence of $R$.  We quantify the error of this approximation and represent it in the abstract Markov model as uncertainty.  We then synthesize an optimal strategy on this model that (i) optimizes the probability of satisfying $\phi$, (ii) is robust against the uncertainty and thus (iii) can be mapped (refined) onto the concrete model $\HybridSystem$. 
In the rest of the paper, we present this solution in detail and show all the proofs in Appendix \ref{sec:proofs}.

\section{Preliminaries}
  \label{sec:preliminaries}
  
\subsection{Markov Models}
\label{sec:markov_models}
We utilize Markov models as abstraction structures.

\begin{mydef}[\mdp] \label{def:mdp}
    A Markov decision process (\mdp) is a tuple $\M = (\Qmdp,\Amdp,\Pmdp,\AP,L)$, where:
    \begin{itemize}
    	\setlength\itemsep{0.5mm}
        \item $\Qmdp$ is a finite set of states,
        \item $\Amdp$ is a finite set of actions,
        \item $\Pmdp: \Qmdp \times \Amdp \times \Qmdp \rightarrow [0,1]$ is a transition probability function.
        \item $\AP$ is a finite set of atomic propositions;
        \item $L: \Qmdp \rightarrow 2^{\AP}$ is a labeling function assigning to each state possibly several elements of $\AP$.
    \end{itemize}
\end{mydef}
The set of actions available at $\qmdp \in \Qmdp$ is denoted by $\Amdp(\qmdp)$. 
The function $\Pmdp$ has the property that $\sum_{\qmdp^\prime \in \Qmdp} \Pmdp(\qmdp,\amdp,\qmdpprime) = 1$ for all pairs $(\qmdp, \amdp)$, where $\qmdp \in \Qmdp$ and $\amdp \in \Amdp(\qmdp)$.

A path $\pathmdp$ through an \mdp is a sequence of states $\pathmdp = \qmdp_0 \xrightarrow{\amdp_0} \qmdp_1 \xrightarrow{\amdp_1} \qmdp_2 \xrightarrow{\amdp_2}  \ldots$ such that $\amdp_i \in \Amdp(\qmdp_i)$ and $\Pmdp(\qmdp_i, \aimdp_i, \allowbreak{\qimdp_{i+1}}) > 0$ for all $i \in \naturals$. We denote 
the last state of a finite path $\pathmdpfin$ by $\last(\pathmdpfin)$ and the set of all finite and infinite paths by $\Pathmdpfin$ and $\Pathmdp$, respectively.

\begin{mydef}[Strategy]
\label{def:strategy}
    A strategy $\str$ of an \mdp model $\M$ is
    a function $\str: \Pathmdpfin \rightarrow \Amdp$ that maps a finite path $\pathmdpfin$ of $\M$ onto an action in $\Amdp$.  If a strategy depends only on $\last(\pathmdpfin)$, it is called a memoryless or stationary strategy.  The set of all strategies is denoted by $\Str$.\footnote{We focus on deterministic strategies as they are sufficient for optimality of \ltl and \sstl properties \cite{abate2008probabilistic,lahijanian2015formal,luna:wafr:2014}.} 
\end{mydef}

\noindent Given a strategy $\str$, a probability measure $\prob$ over the set of all paths (under $\str$) $\Pathmdp$ is induced on the resulting Markov chain \cite{baier2008principles}.


A generalized class of {\mdp}s that allows a range of transition probabilities between states is known as \textit{bounded-pa-\allowbreak{rameter}} \cite{givan2000bounded} or \textit{interval} \mdp (\imdp) \cite{Hahn:QEST:2017}. 

\begin{mydef}[{\imdp}] \label{def:imdp}
    An interval Markov decision process ({\imdp}) is a tuple $\I = (\Qimdp,\Aimdp,\Plow,\Pup,\AP,L)$, where $\Qimdp$, $\Aimdp$, $\AP$, and $L$ are as in Def. \ref{def:mdp}, and
    \begin{itemize}
    	\setlength\itemsep{0.5mm}
        \item $\Plow: \Qimdp \times \Aimdp \times \Qimdp \rightarrow [0,1]$ is a function, where $\Plow(\qimdp,\aimdp,\qimdpprime)$ defines the lower bound of the transition probability from state $\qimdp$ to state $\qimdpprime$ under action $\aimdp \in \Aimdp(\qimdp)$,
        \item $\Pup: \Qimdp \times \Aimdp \times \Qimdp \rightarrow [0,1]$ is a function, where $\Pup(\qimdp,\aimdp,\qimdpprime)$ defines the upper bound of the transition probability from state $\qimdp$ to state $\qimdpprime$ under action $\aimdp \in \Aimdp(\qimdp)$.
    \end{itemize}
\end{mydef}
\noindent
For all $\qimdp,\qimdpprime \in \Qimdp$ and $\aimdp \in \Aimdp(\qimdp)$, it holds that $\Plow(\qimdp,\aimdp,\qimdpprime) \leq \Pup(\qimdp,\aimdp,\qimdpprime)$ and
\begin{equation*}
    \sum_{\qimdpprime \in \Qimdp} \Plow(\qimdp,\aimdp,\qimdpprime) \leq 1 \leq \sum_{\qimdpprime \in \Qimdp} \Pup(\qimdp,\aimdp,\qimdpprime).
\end{equation*}
Let $\probDist(\Qimdp)$ denote the set of discrete probability distributions over $\Qimdp$.  Given $\qimdp \in \Qimdp$ and $\aimdp \in \Aimdp(\qimdp)$, we call $\feasibleDist{\qimdp}{\aimdp} \in \probDist(\Qimdp)$ a \textit{feasible distribution} reachable from $\qimdp$ by $\aimdp$ if 
\begin{equation*}
    \Plow(\qimdp,\aimdp,\qimdpprime) \leq \feasibleDist{\qimdp}{\aimdp}(\qimdpprime) \leq \Pup(\qimdp,\aimdp,\qimdpprime) 
\end{equation*}
for each state $\qimdpprime \in \Qimdp$.
We denote the set of all feasible distributions for state $\qimdp$ and action $\aimdp$ by $\FeasibleDist{\qimdp}{\aimdp}$. 

In {\imdp}s, the notions of paths and strategies are extended from those of {\mdp}s in a straightforward manner.  A distinctive concept instead is that of \textit{adversary}, which is a mechanism that selects feasible distributions from interval sets.\footnote{In the verification literature for {\mdp}s, the notions of strategy, policy, and adversary are often used interchangeably.  The semantics of adversary over {\imdp}s is however distinguished.} 

\begin{mydef}[Adversary]
\label{def:adversary}
    Given an \imdp $\I$, an adversary is a function $\adv: \Pathimdpfin \times \Aimdp \rightarrow \probDist(\Qimdp)$ that, for each  finite path $\pathimdpfin \in \Pathimdpfin$ and action $\aimdp \in \Aimdp(\last(\pathimdpfin))$, assigns a feasible distribution $\adv(\pathimdpfin,\aimdp) \in \FeasibleDist{\last(\pathimdpfin)}{\aimdp}$. 
\end{mydef}

Given a finite path $\pathimdpfin$, a strategy $\str$, and an adversary $\adv$, the semantics of a path of the \imdp is as follows.  At state $\qimdp = \last(\pathimdpfin)$, first an action $\aimdp \in \Aimdp(\qimdp)$ is chosen by strategy $\str$. Then, the adversary $\adv$ resolves the uncertainties and chooses one feasible distribution $\feasibleDist{\qimdp}{\aimdp} \in \FeasibleDist{\qimdp}{\aimdp}$. Finally, the next state $\qimdpprime$ is chosen according to the distribution $\feasibleDist{\qimdp}{\aimdp}$, and the path $\pathimdpfin$ is extended by $\qimdpprime$.

Given a strategy $\str$ and an adversary $\adv$, a probability measure $\prob$ over the set of all paths $\Pathimdp$ (under $\str$ and $\adv$) is induced by the resulting Markov chain \cite{lahijanian2015formal}.

\subsection{Polytopes and their Post Images} \label{sec:polytope}
We use (convex) polytopes as means of discretization in our abstraction.  
Let $m \in \naturals$ and consider the $m$-dimensional Euclidean space $\reals^m$.  A full dimensional (convex) \emph{polytope} $\mathrm{P}$ is defined as the convex hull of at least $m+1$ affinely independent points in $\reals^m$ \cite{grunbaum1967convex}. The \textit{set of vertices} of $\mathrm{P}$ is the set of points $v^\mathrm{P}_1,\ldots,v^\mathrm{P}_{n_\mathrm{P}} \in \reals^m, \, n_\mathrm{P} \geq m+1$, whose convex hull gives $\mathrm{P}$ and with the property that, for any $i = 1,\ldots,n_\mathrm{P}$, point $v^\mathrm{P}_i$  is not in the convex hull of the remaining points $v^\mathrm{P}_1,\ldots,v^\mathrm{P}_{i-1},$ $v^\mathrm{P}_{i+1},\ldots,v^\mathrm{P}_{n_\mathrm{P}}$.  A polytope is completely described by its set of vertices, 
\begin{equation}
    \label{eq:PolyV}
    \mathrm{P} = \conv(v^\mathrm{P}_1,\ldots,v^\mathrm{P}_{n_\mathrm{P}}),
\end{equation}
where $\conv$ denotes the convex hull.  Alternatively, $\mathrm{P}$ can be described as the bounded intersection of at least $m+1$ closed half spaces.  In other words, there exists a $k\geq m+1$, $h_i \in \reals^m,$ and $l_i \in \mathbb{R}, \; i=1,\ldots,k$
such that
\begin{equation}
    \label{eq:PolyH}
    \mathrm{P} = \{x\in \mathbb{R}^m \mid h_i^T x \leq l_i, \, i=1,\ldots,k \}.   
\end{equation}
The above definition can be written as the matrix inequality $H x \leq L$, where $H\in \mathbb{R}^{k\times m}$ and $L \in \mathbb{R}^k$.

Given a matrix $\transfMatrix \in \reals^{m \times m}$, the post image of polytope $\mathrm{P}$ by $\transfMatrix$ is defined as \cite{lahijanian2015formal}:
\begin{equation*}
    \Post(\mathrm{P},\transfMatrix) = \set{\transfMatrix x \mid x \in \mathrm{P}}.
\end{equation*}
This post image is a polytope itself under the linear transformation $\transfMatrix$ 
and can be computed as:
\begin{equation*}
    \Post(\mathrm{P},\transfMatrix) = \conv \big(\{\transfMatrix v^\mathrm{P}_i \mid 1\leq i\leq n_\mathrm{P} \} \big).
\end{equation*}

\section{SHS Abstraction as an IMDP}
  \label{sec:abstraction}



As the first step to approach Problem~\ref{Problem 1}, we abstract the \shs $\HybridSystem$ to an \imdp $\I = (\Qimdp, \Aimdp, \Plow, \allowbreak \Pup, \bar{\AP}, L)$.  Below we overview the construction of the abstraction, and in Sec. \ref{sec:discretization-computation}, we detail the computations involved. 

\textbf{IMDP States.}
We perform a discretization of the hybrid state space $A \times X$.
For each discrete mode $a \in \U$, we partition the corresponding set of interest $X$ into a set of cells (regions) that are non-overlapping, except for trivial sets of measure zero (their boundaries). We assume that each region is a bounded polytope.
We denote by ${\Qimdp}^a=\{\qimdp_1^a,...,\qimdp_{|\Qimdp^a|}^a \}$ the resulting set of regions in mode $a$. To each cell $\qimdp_i^a$, we associate a state of the \imdp $\I$.  We overload the notation by using $\qimdp_i^a$ for both a region in $X$, and a state of $\I$, i.e., $\qimdp_i^a \in \Qimdp$. 
Therefore, the set $(\U \times X) \subset S$ can be represented by $\bar\Qimdp = \bigcup_{a \in \U} \Qimdp^a$. The set of \imdp states is $\Qimdp=\bar \Qimdp \cup \{ \qimdp_{\unsafe} \}$ with $\qimdp_\unsafe$ representing $S \setminus (\U \times X)$, namely 
the complement of $\U \times X$.  



\textbf{IMDP Actions and Transition Probabilities.}
We define the set of actions of $\I$ to be the set of modes $\U$ of $\H$, 
and allow all actions to be available in each state of $\I$, i.e., $\Aimdp(\qimdp) = \Aimdp$ for all $\qimdp \in \Qimdp$.
We define the one-step transition probability from a continuous state $x \in X$ to region $\qimdp \in \bar\Qimdp$ under action (mode) $a \in \Aimdp$ to be defined by the transition kernel $\discreteKernel(\qimdp \mid x,a)$ in \eqref{Eq:TransitionKernel}.
The caveat is that the states of $\I$ correspond to regions in $\H$, and there are uncountably many possible (continuous) initial states (here $x$) in each region, resulting in a range of feasible transition probabilities to the region $q$. Therefore, the transition probability from one region to another can be characterized by a range given by the $\min$ and $\max$ of \eqref{Eq:TransitionKernel} over all the possible points $x$ in the starting region. 
Thus, we can now bound the feasible transition probabilities from state $\qimdp_i \in \bar \Qimdp$ to state $\qimdp_j \in \bar \Qimdp$ from below by
\begin{align}
    \feasibleDist{\qimdp_i}{\aimdp}(\qimdp_j)\geq 
    \min_{x \in q_i} \discreteKernel(q_j \mid x,a),   
\end{align}
and from above by 
\begin{align}
    \feasibleDist{\qimdp_i}{\aimdp}(\qimdp_j)\leq 
    \max_{x \in q_i} \discreteKernel(q_j \mid x,a). 
\end{align}
Thus, for $q_i,q_j \in \bar\Qimdp$, we can define the extrema $\Plow$ and $\Pup$ of the transition probability of $\I$ according to these bounds. 

Similarly, we define the bounds of the feasible transition probabilities to states outside $X$ as
\begin{align}
    \feasibleDist{\qimdp_i}{\aimdp}(\qimdp_u) &\geq 1 - \max_{x \in \qimdp_i} \, \discreteKernel(X  \mid x,a), \label{eq:lowerBound_unsafe} \\
    \feasibleDist{\qimdp_i}{\aimdp}(\qimdp_u) &\leq 1 - \min_{x \in \qimdp_i} \,  \discreteKernel(X  \mid x,a), \label{eq:upperBound_unsafe}
\end{align}
and consequently set the bounds in $\I$ to be
\begin{align}
    \Plow(\qimdp_i,\aimdp,\qimdp_u) &=  1 - \max_{x \in \qimdp_i} \, \discreteKernel(X  \mid x,a),  \label{eq:absLowerBound_unsafe} \\
    \Pup(\qimdp_i,\aimdp,\qimdp_u) &=  1 - \min_{x \in \qimdp_i} \,  \discreteKernel(X  \mid x,a),  \label{eq:absUpperBound_unsafe}
\end{align}
for all $\aimdp \in \Aimdp$ and $\qimdp_i \in \bar \Qimdp$. Finally, since we are not interested in the behavior of $\HybridSystem$ outside of $\U \times X$, we render the state $\qimdp_\unsafe$ of $\I$ absorbing, i.e.,
$\Plow(\qimdp_u,\aimdp,\qimdp_u) = \Pup(\qimdp_u,\aimdp,\qimdp_u) = 1, \, \forall \aimdp \in \Aimdp$. 

\textbf{IMDP Atomic Propositions $\&$ Labels.}
In order to ensure a correct abstraction of $\H$ by $\I$ with respect to the labels of $\H$ and the set $R = \set{r_1,\ldots,r_n}$, even for discretizations of $\U \times X$ that do not respect the regions in $R$, we represent (possibly conservatively) each $r_i$ as well as its complement relative to $X$ through the labeling of the states of $\I$.  Let \[r_{n+i} = X \setminus r_i\] be the complement region of $r_i$ with respect to $X$.  We associate to each $r_{n+i}$ a new atomic proposition $p_{n+i}$ for $1 \leq i \leq n$. 
Intuitively, $p_{n+i}$ represents $\neg p_i$ with respect to $X$.  We define the set of atomic propositions for $\I$ to be
\begin{equation}
	\label{eq:imdp_AP}
	\bar{\AP} = \AP \cup \set{p_{n+1},\ldots,p_{2n}}.
\end{equation}
Then, we design $L: \Qimdp \to 2^{\bar\AP}$ of $\I$ such that
\begin{equation}
	\label{eq:imdp_L}
	p_i \in L(\qimdp) \quad \Leftrightarrow \quad q \subseteq r_i,
\end{equation}
for all $\qimdp \in \bar{\Qimdp}$ and $0 \leq i \leq 2n$, and $L(\qimdp_u) = \emptyset$.  

With this modeling, we capture (possibly conservatively) all the property regions of $\HybridSystem$ by the state labels of $\I$.  Then, a formula $\phi$ over $\AP$ of $\HybridSystem$ can be easily translated to a formula $\bar{\phi}$ on $\bar{\AP}$ of $\I$ by replacing $\neg p_i$ with $p_{n+i}$.  Through this translation, it holds that all the traces that satisfy $\bar{\phi}$ also satisfy $\phi$ and vice versa.  
\begin{remark}
    \label{re:exact_discretization}
    The extension of the atomic propositions in \eqref{eq:imdp_AP} is not necessary if the discretization of $\U \times X$ respects all the regions in $R$, i.e., $\exists Q_{r} \subseteq Q \text{ s.t. } \cup_{q \in Q_r} q = r$ for all $r \in R$.
\end{remark}

\section{Computation of the IMDP}
	\label{sec:discretization-computation}

In this section, we introduce an efficient and scalable method for space discretization and computation for
\begin{align}
\label{timeDiscrBoundsMaxGen}
    \min_{x\in q_i}\discreteKernel(q_j\mid x,a),\qquad \max_{x\in q_i}\discreteKernel(q_j\mid x,a). 
\end{align}
To this end, we first define a \textit{hyper-rectangle} and \textit{proper tranformation function} as follows.
\begin{mydef}[Hyper-rectangle]
    \label{def:hyper-rectangle}
    A \textit{hyper-rectangle} in $\reals^m$ is an $m$-dimensional rectangle defined by the intervals 
    \begin{equation}
        \label{eq:hyper-rectangle}
        [v^{(1)}_l , v^{(1)}_u] \times [v^{(2)}_l , v^{(2)}_u] \times \cdots \times [v^{(m)}_l , v^{(m)}_u],
    \end{equation}
    where vectors $v_l , v_u \in \reals^m$ capture the lower and upper values of the vertices of the rectangle in each dimension, and $v^{(i)}$ denotes the $i$-th component of vector $v$.
\end{mydef}
\begin{mydef}[Proper transformation]
    \label{def:proper-transformation}
    For a polytope $q \subset \reals^m$, the transformation function $\transfMatrix \in \reals^{m \times m}$ is \textit{proper} if $\Post(q,\transfMatrix)$ is a hyper-rectangle.
\end{mydef}
\noindent
We also note that process $\pX$ in mode $a$ is Gaussian with one-step covariance matrix
\begin{equation}
    \label{eq:x-covarian}
    \cov_{\pX}(a) = G(a)^T \cov_w G(a).
\end{equation}
Then, we can characterize $\discreteKernel(q \mid x, a)$ analytically as follows.

\begin{proposition}
    \label{proposition:discrete-kernel-hyperbox}
    For process $\pX$ in mode $a \in \U$, let $\transfMatrix_a = \Lambda_a^{-\frac{1}{2}} V_a^T$ be a transformation function (matrix), where $\Lambda_a = V_a^T \cov_\pX(a) V_a$ is a diagonal matrix whose entries are eigenvalues of $\cov_\pX(a)$ and $V_a$ is the corresponding orthonormal (eigenvector) matrix.
    For a polytopic region $q \subset \reals^m$,
    if $\transfMatrix_a$ is proper,
    then it holds that
    \begin{equation}
        \label{eq:discretekernel_erf_func}
       \discreteKernel(q  \mid x,a) =
        \frac{1}{2^m}\prod_{i=1}^m \Big( \erf(\frac{y^{(i)} -v^{(i)}_l}{\sqrt{2}}) - \erf(\frac{y^{(i)} -v^{(i)}_u}{\sqrt{2}}) \Big),
    \end{equation}
    where $\erf(\cdot)$ is the error function, and $y^{(i)}$ is the $i$-th component of vector $y = \transfMatrix_a \, F(a)x$, and $v^{(i)}_l$, $v^{(i)}_u$ are as in \eqref{eq:hyper-rectangle}.
\end{proposition}
\

A direct consequence of Proposition \ref{proposition:discrete-kernel-hyperbox} is that the optimizations in \eqref{timeDiscrBoundsMaxGen} can be performed on \eqref{eq:discretekernel_erf_func} through a proper transformation, 
as stated by the following corollary.

\begin{corollary}
    \label{corollary:opt_disckernel_transformed}
    For polytopic regions $q_i,q_j \subset \reals^m$ and process $\pX$ in mode $a$, assume 
    $\transfMatrix_a$ is a proper transformation function with respect to $q_j$,
    and define  $q_i'=\Post(q_i, {F(a)})$ and 
    \begin{align}\label{eqn:err_function} f(y)=\frac{1}{2^m}\prod_{i=1}^m \Big( \erf(\frac{y^{(i)} -v^{(i)}_l}{\sqrt{2}}) - \erf(\frac{y^{(i)} -v^{(i)}_u}{\sqrt{2}}) \Big),\end{align}
    where $v_l$ and $v_u$ are as in $\eqref{eq:hyper-rectangle}$.
    Then, it holds that 
    \begin{eqnarray*}
        \min_{x \in q_i}\discreteKernel(q_j  \mid x,a) &=& \min_{y \in \Post(q_i', \transfMatrix_a)} f(y), \\
        \max_{x \in q_i}\discreteKernel(q_j  \mid x,a) &=& \max_{y \in \Post(q_i', \transfMatrix_a)} f(y).
    \end{eqnarray*}
\end{corollary}

The above proposition and corollary show that, 
for a particular proper transformation function $\transfMatrix_a$, 
an analytical form can be obtained for the discrete kernel of the $\imdp$.  This is an important observation because it enables efficient computation for the $\min$ and $\max$ values of the kernel.  Therefore, we use a space discretization to satisfy the condition in Proposition \ref{proposition:discrete-kernel-hyperbox} as described below.  

\subsection{Space Discretization}
\label{sec:space-disc}

For each mode $a \in \Aimdp$, we define the linear transformation function (matrix) of
\begin{equation}
    \label{eq:transfMatrix}
    \transfMatrix_a = \Lambda_a^{-\frac{1}{2}} V_a^T,
\end{equation}
where $\Lambda_a =  V_a^T  \cov_\pX(a)  V_a$ is a diagonal matrix whose entries are the eigenvalues of $\cov_\pX(a)$, and $V_a$ is the corresponding orthonormal (eigenvector) matrix. The discretization of the continuous set $X$ in mode $a$ is achieved by using a grid in the transformed space by $\transfMatrix_a $.  That is, we first transform $X$ by $\transfMatrix_a$, and then discretize it using a grid.
This method of discretization guarantees that, for each $q^a \in Q^a$, $\Post(q^a , \transfMatrix_a)$ is a hyper-rectangle, i.e., $\transfMatrix_a$ is proper.  Hence, we can use the result of Proposition~\ref{proposition:discrete-kernel-hyperbox} and Corollary~\ref{corollary:opt_disckernel_transformed} for the computation of the values in~\eqref{timeDiscrBoundsMaxGen}.  


\begin{remark}
    \label{re:discretization-X}
    For an arbitrary geometry of $X$, it may not be possible to obtain a discretization such that $\bigcup_{q^a\in Q^a} q^a = X$.  Nevertheless, by using a discretization that under-approxima\allowbreak{tes} $X$, i.e., $\bigcup_{q^a\in Q^a} q^a \subseteq X$, in each mode $a$, we can compute a lower bound on the probability of satisfaction of a given property $\phi$. For a better approximation, the grid can be non-uniform, allowing in particular for smaller cells near the boundary of $X$, as in \cite{esmaeil2013adaptive}.
\end{remark}

\subsection{Transition Probability Bounds}
\label{sec:computation-bounds}

We distinguish between transitions from $q \in \bar\Qimdp$ to the states in $\bar{\Qimdp}$ and to the state $\qimdp_\unsafe$.

\subsubsection{Transitions to $q \in \bar{Q}$}
We present two approaches to solving the values for \eqref{timeDiscrBoundsMaxGen}.  The first approach is based on \textit{Karush-Kuhn-Tucker} (\kkt) conditions \cite{bertsekas2014constrained}, which sheds light into the optimization problem and lays down the conditions on where to look for the optimal points, giving geometric intuition.  This method boils down to solving systems of non-linear equations, which turns out to be efficient and exact for low-dimensional systems.  In the second approach, we show that the problem reduces  to a convex optimization problem, allowing the adoption of existing optimization tools and hence making the approach suitable for high-dimensional systems. 

\textbf{KKT Optimization Approach:}
In the next theorem, we use the result of Corollary \ref{corollary:opt_disckernel_transformed} and the \kkt conditions \cite{bertsekas2014constrained} to compute the exact values for \eqref{timeDiscrBoundsMaxGen}. 
\begin{theorem}
    \label{th:SpaceDiscretizationKKTOnditions}
    For polytopic regions $q_i,q_j \subset \reals^m$ and proper transformation matrix $\transfMatrix_{a}$ with respect to $q_j$, let
    \begin{equation*}
        \Post(q'_i , \transfMatrix_{a} ) = \{ y\in \mathbb{R}^m \mid H y \leq b \},
    \end{equation*}
    where $q'_i = \Post(q_i , F(a))$, $H \in \reals^{k \times m}$, $b \in \reals^{m}$, and $k \geq m+1$, and 
    introduce the following conditions:
    \begin{itemize}
        \setlength\itemsep{0.5mm}
        \item {$\mathbf{Condition \,1}$}: $y$ is at the center of $\Post(q_j , \transfMatrix_{a} )$, i.e.,  $y=(\frac{v^{(1)}_u +v^{(1)}_l}{2}, \ldots,\frac{v^{(m)}_u+v^{(m)}_l}{2})$.
        \item {$\mathbf{Condition\, 2}$}: $y$ is a  vertex of  $\Post(q_i' ,\transfMatrix_{a} ).$ 
        \item {$\mathbf{Condition\, 3}$}: $y$ is on the boundary of $\Post(q_i' ,\transfMatrix_{a} )$, where $r \geq 1$  of the $k$ half-spaces that define $\Post(q_i' ,\transfMatrix_{a} )$ intersect, and 
        \begin{align*}
            \nabla f(y) = \bar{H}^T \mu,
        \end{align*}
        for vector $\mu=(\mu_1,\ldots,\mu_r)$ of non-negative constants, and submatrix $\bar{H} \in \mathbb{R}^{r\times m}$ that contains only the rows of $H$ that correspond to the $r$-intersecting half-spaces at $y$.
        \item {$\mathbf{Condition\, 4}$}: $y$ is as in Condition $3$, and
        \begin{equation*}
            \nabla f(y) = - \bar{H}^T \mu,
        \end{equation*}
        for vector $\mu=(\mu_1,\ldots,\mu_r)$ of non-negative constants, and $\bar{H}$ is defined as in Condition $3$.
    \end{itemize}
    Then, it follows that the point $y \in \Post(q_i' , \transfMatrix_{a} )$ that satisfies Condition $1$ necessarily maximizes $f(y)$. If Condition $1$ cannot be satisfied, then the maximum is necessarily given by one of the points that satisfy Condition $2$ or $3$.  Furthermore, the point $y \in \Post(q_i' , \transfMatrix_a)$ that minimizes $f(y)$ necessarily satisfies Condition $2$ or $4.$
\end{theorem}

\noindent

Theorem \ref{th:SpaceDiscretizationKKTOnditions} identifies the arguments (points $y 
\in \Post(q_i' , \transfMatrix_{a} )$) that give rise to the optimal values of $\discreteKernel$ in \eqref{timeDiscrBoundsMaxGen}.  
Then, the actual optimal values of $\discreteKernel$ can be computed by \eqref{eqn:err_function} as guaranteed by Corollary~\ref{corollary:opt_disckernel_transformed}.  Therefore, from Theorem \ref{th:SpaceDiscretizationKKTOnditions}, an algorithm can be constructed to generate a set of finite candidate points based on Conditions 1-4 and to obtain the exact values of \eqref{timeDiscrBoundsMaxGen} by plugging those points into \eqref{eqn:err_function}.

In short, Condition 1 maximizes the unconstrained problem and gives rise to the global maximum.  Hence, if the center of $q_j$ is contained in $\Post(q_i' , \transfMatrix_a)$, no further check is required for maximum.  If not, the maximum is given by a point on the boundary of $\Post(q_i' , \transfMatrix_a)$.  It is either a vertex (Condition 2) or a boundary point that satisfies Condition 3.  The minimum is always given by a boundary point, which can be either a vertex or a boundary point that satisfies Condition 4.  Note that Conditions 3 and 4 are similar and both state that the optimal value of $\discreteKernel$ is given by a point where the gradient of $\discreteKernel$ becomes linearly dependent on the vectors that are defined by the intersecting half-spaces of $\Post(q_i' , \transfMatrix_a)$ at that point.  Each of these two conditions defines a system of $m$ equations and $r < m$ variables, which may have a solution only if some of the equations are linear combinations of the others.


The above algorithm computes the exact values for the transition probability bounds. It is computationally efficient for small dimensional systems, e.g., $m < 4$.  For large $m$, however, the efficiency drops because the number of boundary constraints that need to be checked and solved for in Conditions 3 and 4 increases, in the worst case, exponentially with $m$.  Below, we propose an equivalent but more efficient method to compute $\min$ and $\max$ of $\discreteKernel$ for large dimensional systems, e.g., $m \geq 4$.

\textbf{Convex Optimization Approach:}
In order to show how upper and lower bounds of $f(y)$ can be efficiently computed using convex optimization tools, we need to introduce the definition of \emph{concave} and \emph{log-concave} functions.
\begin{mydef}[Concave Function]
A function $g:\mathbb{R}^m \to \mathbb{R}$ is said to be concave if and only if for  $y_1,y_2 \in \mathbb{R}^m$, $\lambda \in [0,1]$ 
$$ g( \lambda y_1 + (1-\lambda) y_2) \geq \lambda g(y_1) + (1-\lambda) g(y_2).  $$
\end{mydef}

\begin{mydef}[Log-concave Function]
A function $g:\mathbb{R}^m \to \mathbb{R}$ is said to be log-concave if and only if $log(g)$ is a concave function. That is, for $y_1,y_2 \in \mathbb{R}^m$, $\lambda \in [0,1]$ 
$$ g( \lambda y_1 + (1-\lambda) y_2) \geq  g(y_1)^{\lambda} g(y_2)^{(1-\lambda)}.  $$
\end{mydef}
\noindent
In the following proposition, we show that $ f(y)$, as defined in Corollary \ref{corollary:opt_disckernel_transformed}, is log-concave. This enables efficient computation of the upper and lower bounds of $ f(y)$ through standard convex optimization techniques such as gradient descent or semidefinite programming \cite{boyd2004convex}. Hence, we can make use of readily available software tools, e.g., NLopt~\cite{johnson2014nlopt}, which have been highly optimized in terms of efficiency  and scalability.  
\begin{proposition}
\label{Proposition-LogConcaveFundaction}
    $ f(y)$, as defined in Corollary \ref{corollary:opt_disckernel_transformed}, is a log-concave function.
\end{proposition}

\subsubsection{Transitions to sink state $\qimdp_\unsafe$}
\label{sec:tran-unsafe}
Here, we focus on the transition probabilities to state $q_u$ in \eqref{eq:absLowerBound_unsafe} and \eqref{eq:absUpperBound_unsafe}.  To this end, we need to compute  
\begin{equation}
    \label{eq:lower-upper-unafe}
	\max_{x \in \qimdp_i} \, \discreteKernel(X \mid x,\aimdp), \quad
	\min_{x \in \qimdp_i} \, \discreteKernel(X \mid x,\aimdp).
\end{equation}
We can efficiently compute bounds for these quantities by using the results obtained above.  The following proposition shows this efficient method of computation.

\begin{proposition}
\label{prop:tran_unsafe}
    Let $\check \Qimdp^a$ and $\hat \Qimdp^a$ be two sets of polytopic regions in mode $a$ such that 
    \begin{equation*}
    	\bigcup_{q\in \check \Qimdp^a} q \; \subseteq \; X \; \subseteq \; \bigcup_{q \in \hat \Qimdp^a} q,
    \end{equation*}
    and $\transfMatrix_{a}$ be a proper transformation function 
    for every $q \in \check \Qimdp^a\cup \hat{\Qimdp}^a$, and call
    \begin{equation}
    	\label{eq:err_fun_q}
        f(y,q) = \frac{1}{2^m} \prod_{i=1}^m \Big( \erf(\frac{y^{(i)} -v^{(i)}_{l,q}}{\sqrt{2}}) - \erf(\frac{y^{(i)} -v^{(i)}_{u,q}}{\sqrt{2}}) \Big),
    \end{equation}
    where $v_{l,q}$ and $v_{u,q}$ are as in \eqref{eq:hyper-rectangle} for $q$.
    Then, it holds that
    \begin{align}
    	\label{eq:psafe_max_bound}
    	\max_{x \in \qimdp_i} \, \discreteKernel(X \mid x,\aimdp) &\leq \max_{y \in \Post(q_i', \transfMatrix_a)} \sum_{q \in \hat \Qimdp^a}  f(y,q), \\
    	\label{eq:psafe_min_bound}
    	\min_{x \in \qimdp_i} \, \discreteKernel(X \mid x,\aimdp) &\geq  \min_{y \in \Post(q_i', \transfMatrix_a)} \sum_{q \in \Qimdp^a}  f(y,q),
    \end{align}
    where $q_i'=\Post(q_i , F(a))$.
\end{proposition}

\noindent

Intuitively, Proposition \ref{prop:tran_unsafe} states that, with a particular choice of discretization, i.e., a grid in the transformed space, the transition probability to $X$ is equal to the sum of the transition probabilities to the discrete regions, where each discrete transition kernel is given by the close-form function $f(y,q)$ in \eqref{eq:err_fun_q}.  If $X$ cannot be precisely discretized with a grid (in the transformed space), then the upper and lower bounds of the transition probabilities are given by the over- and under-approximating grids ($\hat \Qimdp^a$ and $\check \Qimdp^a$), respectively. 


\begin{remark}
    For the computation of the values in \eqref{eq:psafe_max_bound} and \eqref{eq:psafe_min_bound}, Proposition \ref{Proposition-LogConcaveFundaction} can be applied, making both methods of \kkt and convex optimization applicable. 
\end{remark}

\section{Strategy Synthesis as a Game}
  \label{sec:synthesis}


Recall that our objective is, given the compact set $X$ and a \sstl or \ltl formula $\phi$, to compute a strategy for $\H$ that maximizes the probability of satisfying $\phi$ without exiting $X$.  The \imdp abstraction $\I$, as constructed above, captures (possibly conservatively) the behavior of the \shs $\H$ with respect to the regions of interest $R$ within $X$, and the probabilities of exiting $X$ are encompassed via the state $\qimdp_\unsafe$. Since state $\qimdp_\unsafe$ is absorbing, the paths of $\I$ are not allowed to exit and re-enter $X$; as such, the analysis on $\I$ narrows the focus to dynamics within set $X$, as desired.  Therefore, we can focus on finding a strategy for $\I$ that is robust against all the uncertainties (errors) introduced by the discretization of $\U \times X$ and which maximizes $\phi$. 

The uncertainties in $\I$ can be viewed as the nondeterministic choice of a feasible transition probability from one \imdp state to another under a given action.  Therefore, we interpret a synthesis task over the \imdp as a 2-player stochastic game, where Player 1 chooses an action $\aimdp \in \Aimdp$ at state $\qimdp \in \Qimdp$, and Player 2 chooses a feasible transition probability distribution $\feasibleDist{\qimdp}{\aimdp} \in \FeasibleDist{\qimdp}{\aimdp}$.  Towards robust analysis, we set up this game as adversarial: the objectives of Players 1 and 2 are to maximize and minimize the probability of satisfying $\phi$, respectively.  Hence, the goal becomes to synthesize a strategy for Player 1 that is robust against all adversarial choices of Player 2 and maximizes the probability of achieving $\phi$.

In order to compute this strategy, we first translate $\phi$ over $\AP$ into its equivalent formula $\bar\phi$ over $\bar\AP$.  Then, we construct a \textit{deterministic finite automaton} (\dfa) $\Adfa_{\bar\phi}$ that precisely accepts all the good prefixes that satisfy $\bar\phi$ \cite{kupferman:FMSD:2001}.  
\begin{mydef}[\dfa]
	\label{def:dfa}
	A \dfa constructed from a \ltl or \sstl formula $\bar\phi$ is a tuple $\Adfa_{\bar\phi} = (\Zdfa, 2^{\bar\AP}, \Trandfa, \zdfa_0, \Fdfa)$, where
	$\Zdfa$ is a finite set of states, 
	$2^{\bar\AP}$ is the set of input alphabets,
	$\Trandfa:\Zdfa \times 2^{\bar \AP} \to \Zdfa$ is the transition function,
	$\zdfa_0\in \Zdfa$ is the initial state, and
	$\Fdfa \subseteq \Zdfa$ is the set of accepting states.
\end{mydef}
\noindent
A \textit{finite run} of $\Adfa_{\bar \phi}$ on a trace $\trace = \trace_1 \cdots \trace_n$, where $\trace_i \in 2^{\bar\AP}$, is a sequence of states $\mu = z_0 z_1 \ldots z_n$ with $z_i = \Trandfa(z_{i-1},\trace_i)$ for $i = 1,\ldots,n$.  Run $\mu$ is called \textit{accepting} if $\mu_n \in \Fdfa$. Trace $\trace \models \bar\phi$ iff its corresponding run $\mu$ in $\Adfa_{\bar \phi}$ is accepting. 

Next, we construct the product \imdp $\I_{\bar\phi} = \I \times \Adfa_{\bar\phi}$, which is a tuple $\I_{\bar\phi} = (\Qimdp_{\bar\phi}, \Aimdp_{\bar\phi}, \Plow_{\bar\phi}, \Pup_{\bar\phi}, \Fimdp)$, where
\begin{gather*}
	\Qimdp_{\bar\phi} = \Qimdp \times \Zdfa, \quad \Aimdp_{\bar\phi} = \Aimdp, \quad \Fimdp = \Qimdp \times \Fdfa, \\
	\Plow_{\bar\phi}((\qimdp,\zdfa),\aimdp,(\qimdp',\zdfa')) = 
		\begin{cases}
			\Plow(\qimdp,\aimdp,\qimdp') & \text{if } \zdfa' = \Trandfa(z,L(\qimdp'))\\
			0 & \text{otherwise},
		\end{cases} \\
	\Pup_{\bar\phi}((\qimdp,\zdfa),\aimdp,(\qimdp',\zdfa')) = 
		\begin{cases}
			\Pup(\qimdp,\aimdp,\qimdp') & \text{if } \zdfa' = \Trandfa(z,L(\qimdp'))\\
			0 & \text{otherwise},
		\end{cases}
\end{gather*}
for all $\qimdp,\qimdp' \in \Qimdp$, $\aimdp \in \Aimdp$, and $\zdfa \in \Zdfa$.  Intuitively, $\I_{\bar\phi}$ contains both $\I$ and $\Adfa_{\bar\phi}$ and hence can identify all the paths of $\I$ that satisfy $\bar\phi$, i.e., the satisfying paths terminate in $\Fimdp$ since their corresponding $\Adfa_{\bar\phi}$ runs are accepting.  Therefore, the synthesis problem reduces to computing a robust strategy on $\I_{\bar\phi}$ that maximizes the probability of reaching $\Fimdp$.  This problem is equivalent to solving the \textit{maximal reachability probability problem} \cite{wu2008reachability,lahijanian2015formal} as explained below.  

Given a strategy $\str$ on an \imdp, the probability of reaching a terminal state from each state is necessarily a range for all the available adversarial choices of Player 2.  Let $\plow^\str(\qimdp)$ and $\pup^\str(\qimdp)$ denote lower and upper bounds for the probability of reaching a state in $\Fimdp$ starting from $\qimdp \in \Qimdp_{\bar\phi}$ under $\str$.  
Derived from the Bellman equation, we can compute the optimal lower bound by recursive evaluations of
\begin{align}
	\label{eq:bellman_lower}
	\hspace{-2mm}
	\plow^{\str^*}\hspace{-1mm}(\qimdp) = 
		\begin{cases}
			1 & \text{if } \qimdp \in \Fimdp \\
			\hspace{-1mm}\max\limits_{\aimdp \in \Aimdp(\qimdp)} \min\limits_{\feasibleDist{\qimdp}{\aimdp} \in \FeasibleDist{\qimdp}{\aimdp} } 
			\sum\limits_{\qimdp' \in \Qimdp_{\bar\phi}} \feasibleDist{\qimdp}{\aimdp}(\qimdp') \plow^{\str^*}\hspace{-1mm}(\qimdp') & \text{otherwise,}
		\end{cases}
\end{align}
for all $\qimdp \in \Qimdp_{\bar\phi}$.  Each iteration of this Bellman equation involves a minimization over the adversarial choices, which can be computed through an ordering of the states of $\I_{\bar\phi}$ \cite{givan2000bounded,lahijanian2015formal}, and a maximization over the actions.  
This Bellman equation is guaranteed to converge in finite time \cite{lahijanian2015formal,wu2008reachability} and results in the lower-bound probability $\plow^{\str^*}\hspace{-1mm}(\qimdp)$ for each $\qimdp \in \Qimdp_{\bar\phi}$ and in a stationary (memoryless) strategy $\str^*$.  The upper bounds are similarly given by recursive evaluations of 
\begin{align}
	\label{eq:bellman_upper}
	\pup^{\str^*}\hspace{-1mm}(\qimdp) = 
		\begin{cases}
			1 & \text{if } \qimdp \in \Fimdp \\
			\max\limits_{\feasibleDist{\qimdp}{\str^*} \in \FeasibleDist{\qimdp}{\str^*} } 
			\sum\limits_{\qimdp' \in \Qimdp_{\bar\phi}} \feasibleDist{\qimdp}{\str^*}(\qimdp') \pup^{\str^*}\hspace{-1mm}(\qimdp') & \text{otherwise,}
		\end{cases}
\end{align}
which is also guaranteed to converge in finite time. 

The optimal strategy $\str^*$ on $\I_{\bar\phi}$ can be mapped onto the states and actions of the abstraction \imdp $\I$, resulting in a (history-dependent) strategy.  By construction, then the optimal lower and upper probability bounds of satisfying $\phi$ from the states of $\I$ are:
\begin{equation}
	\label{eq:prob-bounds-phi}
	\plow^{\str^*}_{\phi}(q) = \plow^{\str^*}\hspace{-1mm}((q,z_0)), \quad \pup^{\str^*}_{\phi}(q) = \pup^{\str^*}\hspace{-1mm}((q,z_0)),
\end{equation}
for all $\qimdp \in \Qimdp$ of $\I$.

The complexity of the above strategy synthesis algorithm is polynomial in the size of the \imdp $\I_{\bar\phi}$ \cite{wu2008reachability,lahijanian2015formal}   
and exponential in the size of the formula $\phi$ (in the worst case) \cite{kupferman:FMSD:2001}.  
Note that the size of $\phi$ used to express the properties of \shs is typically small.

\section{Correctness}
  \label{sec:correctness}
 

We show that the strategy $\str^*$ computed over $\I$ can be refined over (mapped onto) $\H$ and the lower probability bound $\plow^{\str^*}_{\phi}$ on $\I$ always holds for the hybrid system $\H$.  The upper bound $\pup^{\str^*}_{\phi}$ also holds for $\H$ if the discretization respects the regions in $R$.
In the case that the discretization is not $R$-respecting, a modified upper bound that holds for $\H$ can be computed with a small additional step as detailed below.

Let $\L: S \rightarrow \Qimdp$ be a function that maps the hybrid states $s \in S$ to their corresponding discrete regions (states of $\I$), i.e, $\L(s) = \qimdp \in \Qimdp$ if $s \in \qimdp$.  With a slight abuse of notations, we also use $\L$ to denote the mapping from the finite paths of $\H$ to their corresponding paths of $\I$, i.e.,
\begin{equation*}
	\pathH{k} = s_0 s_1 \ldots s_k  \quad \Rightarrow \quad \L(\pathH{k}) = \L(s_0)\L(s_1) \ldots \L(s_k).
\end{equation*}
Then, the \imdp strategy $\str^*$ correctly maps to a switching strategy $\strH^*$ for $\H$ via 
\begin{equation}
	\label{eq:strategy_mapping}
	\strH^*(\pathH{k }) = \str^*(\L(\pathH{k })).
\end{equation}



The following theorem shows that for a given $\phi$, the  probability bounds $\plow^{\str^*}_\phi$ and $\pup^{\str^*}_\phi$ are guaranteed to hold for the  process $\pH$  under $\strH^*$ as constructed above.

\begin{theorem}
	\label{th:Correctness}
	Given a \shs $\HybridSystem$, a continuous set $X$, and a \ltl or \sstl formula $\phi$, let $\I$ be the \imdp abstraction of $\HybridSystem$ as described in Section \ref{sec:abstraction} through a discretization that respects the regions of interest in $R$.  Further, let $\str^*$ be the strategy on $\I$ computed by \eqref{eq:bellman_lower} and \eqref{eq:bellman_upper} with probability bounds $\plow^{\str^*}_\phi$ and $\pup^{\str^*}_\phi$ in \eqref{eq:prob-bounds-phi}. Refine $\str^*$ into a switching strategy $\strH^*$ as in \eqref{eq:strategy_mapping}. Then, for any initial hybrid state $s_0 \in S$, where $s_0 \in q_0 \in \Qimdp$, it holds that 
	\begin{equation}
		\label{eq:hybridsys_safety_bounds}
	P(\phi \mid s_0, X, \strH^*) \in \big[\plow^{\str^*}_\phi(q_0), \, \pup^{\str^*}_\phi(q_0) \big].
	\end{equation}
\end{theorem}

Note that an assumption in Theorem \ref{th:Correctness} is that the discretization $\Qimdp$ respects the regions in $R$.  If this assumption is violated, then the lower bound $\plow^{\str^*}_\phi$ still holds, unlike the upper bound $\pup^{\str^*}_\phi$.  That is because we design the labeling function $L$ of $\I$ to under-approximate the regions of interest $r \in R$, making the upper bound $\pup^{\str^*}_\phi$ valid with respect to the under-approximate representation of $R$ by $L$ but possibly under-approximated with respect to the actual $R$.  To compute an upper bound that accounts for this, we need to design a new labeling function that over-approximates the labels of each region, as follows.  Let $L': \Qimdp \to \bar{\AP}$ be this labeling function with
\begin{equation}
	\label{eq:label-over-approximating}
	p_i \in L'(q) \quad \Leftrightarrow \quad \exists (a,x) \in q \text{ s.t. } x \in r_i,
\end{equation}
where $p_i \in \bar \AP$ is the associated proposition to $r_i \in R$.  Then, we can compute the over-approximated upper bound $\pup^{\prime\str^*}_\phi$ via \eqref{eq:bellman_upper} on the product \imdp $\I^\prime_{\bar\phi}$ constructed using $L'$. 

\begin{lemma}
	\label{prop:upper_bound}
	If abstraction $\I$ is constructed through a discretization that does not respect the regions in $R$, then
	\begin{equation}
		\label{eq:hybridsys_safety_lowerbound}
	P(\phi \mid s_0, X, \strH^*) \in \big[ \plow^{\str^*}_\phi(q_0), \pup^{\prime \str^*}_\phi(q_0) \big],
	\end{equation}
	where $\pup^{\prime \str^*}_\phi$ is computed via \eqref{eq:bellman_upper} using the labels in \eqref{eq:label-over-approximating}.
\end{lemma}

Theorem \ref{th:Correctness} and Lemma \ref{prop:upper_bound} guarantee that the satisfaction probability of $\phi$ for the process $\pH$, solution of the \shs $\HybridSystem$, is contained in the probability interval computed on the abstraction $\I$.  The size of this interval depends on the difference of the one-step transition probability bounds of $\Plow$ and $\Pup$ as well as the embedded approximations in the labeling functions $L$ and $L'$ in $\I$,
which can be viewed as the error induced by space discretization of $\H$ cast into the abstraction $\I$.
This error can be tuned by the size of the discretization: in particular, in the limit of an infinitely fine grid, the error of the abstraction goes to zero, and the \imdp abstraction is refined into an \mdp, namely for all $q,q' \in Q$ and $a \in A(q)$, $\Plow(q,a,q') \rightarrow P(q,a,q') \leftarrow \Pup(q,a,q')$.

\begin{remark}
    In practice, the interest in synthesis problems is typically on deriving lower bounds for the probability, whereas the upper bound computation is useful for error analysis. 
\end{remark}

\begin{remark}
    With a simple modification, the proposed framework can be used for verification of \shs $\H$ against property $\phi$: (i) compute the lower-bound probability by replacing $\max_{a\in A(q)}$ with $\min_{a\in A(q)}$ in \eqref{eq:bellman_upper} on abstraction $\I$ with labeling function $L$, and (ii) compute the upper-bound probability by replacing $\min_{\feasibleDist{\qimdp}{\aimdp} \in \FeasibleDist{\qimdp}{\aimdp}} $ with $\max_{\feasibleDist{\qimdp}{\aimdp} \in \FeasibleDist{\qimdp}{\aimdp}} $ in \eqref{eq:bellman_upper} on abstraction $\I$ with labeling function $L'$. 
\end{remark}

\section{Experimental Results}
  \label{sec:experiments}
 \newcommand{\faust}{\textsc{faust}$^2$\xspace}
\newcommand{\gd}{\textsc{gd}\xspace}
\newcommand{\cpp}{\textsc{c++}\xspace}
\newcommand{\matlab}{\textsc{matlab}\xspace}


We implement the abstraction and synthesis algorithms and test their performance on three case studies.  We first present a two dimensional stochastic process with a single mode and perform a comparison against the algorithms and tool \faust \cite{SA15} in Case Study 1.  Next, we consider a two dimensional, two-mode model and show the synthesis over unbounded-time properties in Case Study 2.   
Last, we analyze the scalability of the proposed techniques over increasing continuous dimension of the \shs in Case Study 3. 

The implementation of the abstraction algorithm is in \matlab and \cpp: more precisely, the approach based on \kkt method is in \matlab (as a proof of concept), and the convex optimization method with \textit{gradient decent} (\gd) is in \cpp.  The synthesis algorithm over the \imdp is also implemented in \cpp.  
The experiments are run on an Intel Core i7-8550U CPU at 1.80GHz $\times$ 8 machine with 8 GB of RAM. 


\subsection{Case Study 1 - Formal Verification}
\label{subsec:Cs1}
We consider a stochastic process with dynamics in~\eqref{eq:switched-dynamics} and a single discrete mode ($A =\{a_1\}$), 
where 
\begin{equation*}
  F(a_1) =  
  \begin{pmatrix}
    0.85 & 0\\
    0    &  0.90 \\
 \end{pmatrix},
 \quad
 G(a_1)= 
 \begin{pmatrix}
    0.15 & 0\\
    0    &  0.05 \\
 \end{pmatrix},
\end{equation*}
with $\safeSet=[-1,1]\times[-1,1]$ and safety property
\[\phi_1=  \mathcal{G}^{\leq k} \safeSet.\]

We compare the verification results of the above model using our method against  those of the state-of-the-art tool \faust \cite{SA15}.  Namely, we compare probability of satisfaction of $\phi_1$, computation times, and errors for a range of values for time horizon $k$ and grid sizes.  
To obtain the \imdp abstraction of our method, we used a uniform grid discretization per Sec. \ref{sec:discretization-computation}. 
Tool \faust abstracts the model into an \mdp and treats the error as a separate parameter.  The grid generated in \faust is based on computation of the global Lipschitz constant via integrals~\cite{SA15}.  We define the error of the \imdp method to be 
    $ \err_q = \pup^*_\phi(q) - \plow^*_\phi(q)   $
for each state, and the global error to be $\err_{max}=\max_{q \in \Qimdp} \err_q$.  Similarly, for \faust the resulting error corresponds to the maximum error over all the states. 
The \faust tool is written in \matlab and run over this platform, however additionally for fair comparison we have re-implemented the abstraction based on \faust in the \cpp language (cf. corresponding lines in Table~\ref{tab:comp_FI}). 

The results are shown in Table~\ref{tab:comp_FI} for $k=2$ and various grid sizes. We saturate conservative errors output by \faust that are greater than $1$ to this value. For the particular grid $|Q|=3722$, the lower bound probabilities of satisfying $\phi_1$ are shown in Fig.~\ref{fig:Cs1_Ex2} within Appendix~\ref{app:Cs1}. 
As evident in Table \ref{tab:comp_FI}, 
our approach greatly outperforms the state of the art. With respect to the error generated for the same grid size, our method has significantly (an order of magnitude) smaller error than \faust. Our \imdp method also   requires lower computation times.
We also note that, as guaranteed by the theory (Theorem \ref{th:SpaceDiscretizationKKTOnditions} and Proposition \ref{Proposition-LogConcaveFundaction}), both \kkt and \gd approaches compute the same error.

\begin{table}[h!]
    \centering
    \resizebox{\columnwidth}{!}{
    \begin{tabular}{l|cccc}
    \textbf{Tool} & \textbf{Impl.} & $|\bar{\mathbf{Q}}|$    & \textbf{Time taken}  & \textbf{Error} \\
    \textbf{Method} & \textbf{Platform} & (states) & (secs)             &   $\err_{\max}$ \\ \hline \hline 
    \imdp (\kkt)    & \matlab    & 361       & 19.789   & 0.211\\
    \imdp (\gd)     & \cpp       & 361       & 29.003  & 0.211 \\  
    \faust 	    & \matlab    & 361       & 108.265  & 1.000\\ 
    \faust          & \cpp       & 361       &  136.71  & 1.000 \\ 
    \hline 

    \imdp (\kkt)    & \matlab    & 625       & 145.563  &  0.163 \\  
    \imdp (\gd)     & \cpp       & 625       & 117.741   & 0.163\\ 
    \faust          & \matlab    & 625       & 285.795  & 1.000 \\ 
    \faust      & \cpp       & 625       &  302.900 & 1.000  \\ \hline

    \imdp (\kkt)    & \matlab    & 1444      & 4464.783 & 0.109 \\
    \imdp  (\gd)    & \cpp       & 1444      & 510.920   & 0.109 \\
    \faust      & \matlab    & 1444      &  1445.441 &  1.000\\
    \faust      & \cpp       & 1444      & 1201.950    & 1.000 \\ \hline %

    \imdp (\kkt)    & \matlab    & 2601     & 28127.256   &0.082    \\
    \imdp (\gd)     & \cpp       & 2601     & 2939.050   & 0.082\\
    \faust      & \matlab    & 2601     &5274.578    & 0.995  \\ 
    \faust      & \cpp       & 2601     & 3305.490   & 0.995\\ \hline 

    \imdp (\kkt)    & \matlab    & 3721      &  Time out\footnote{9 hours+ and no solution} & - \\ 
    \imdp (\gd)     & \cpp       & 3721      & 3973.28 &  0.068\\ 
    \faust      & \matlab    & 3721      & 11285.313 & 0.832 \\ 
    \faust      & \cpp       & 3721      & 7537.750  & 0.832 \\ \hline
    \hline
    \end{tabular}    }
   \caption{Comparison of verification results of our \imdp algorithms against \faust for $\phi_1$ with $k = 2$.}
    \label{tab:comp_FI}
\end{table}

\begin{figure}[t]
	\centering
\begin{subfigure}{.49\columnwidth}
     \centering
    \includegraphics[width=.99\columnwidth]{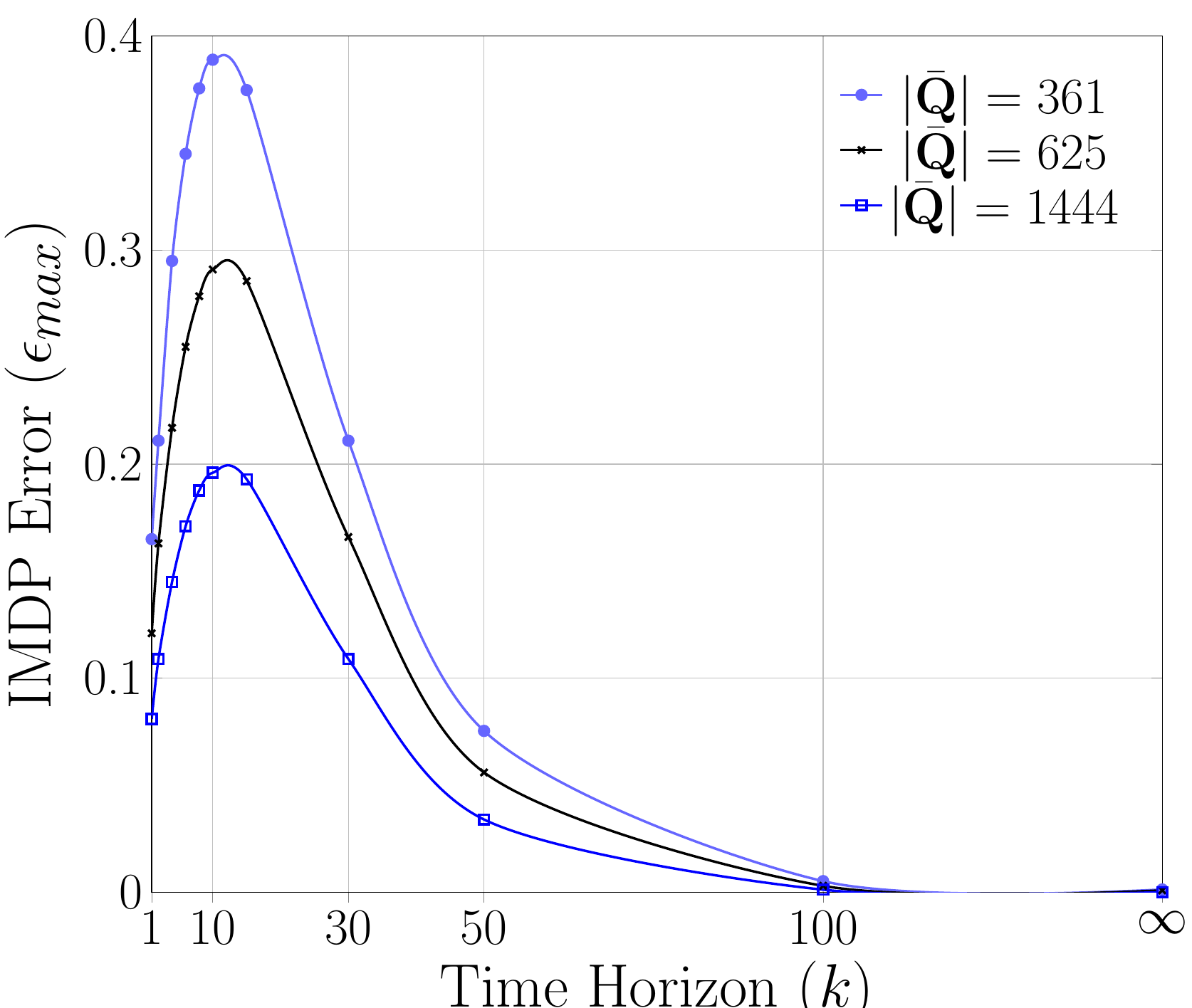}
     \caption{\imdp} 
      \label{fig:E_IMDP}
   \end{subfigure}
  \begin{subfigure}{.49\columnwidth}
  \centering
 \includegraphics[width=1.01\columnwidth]{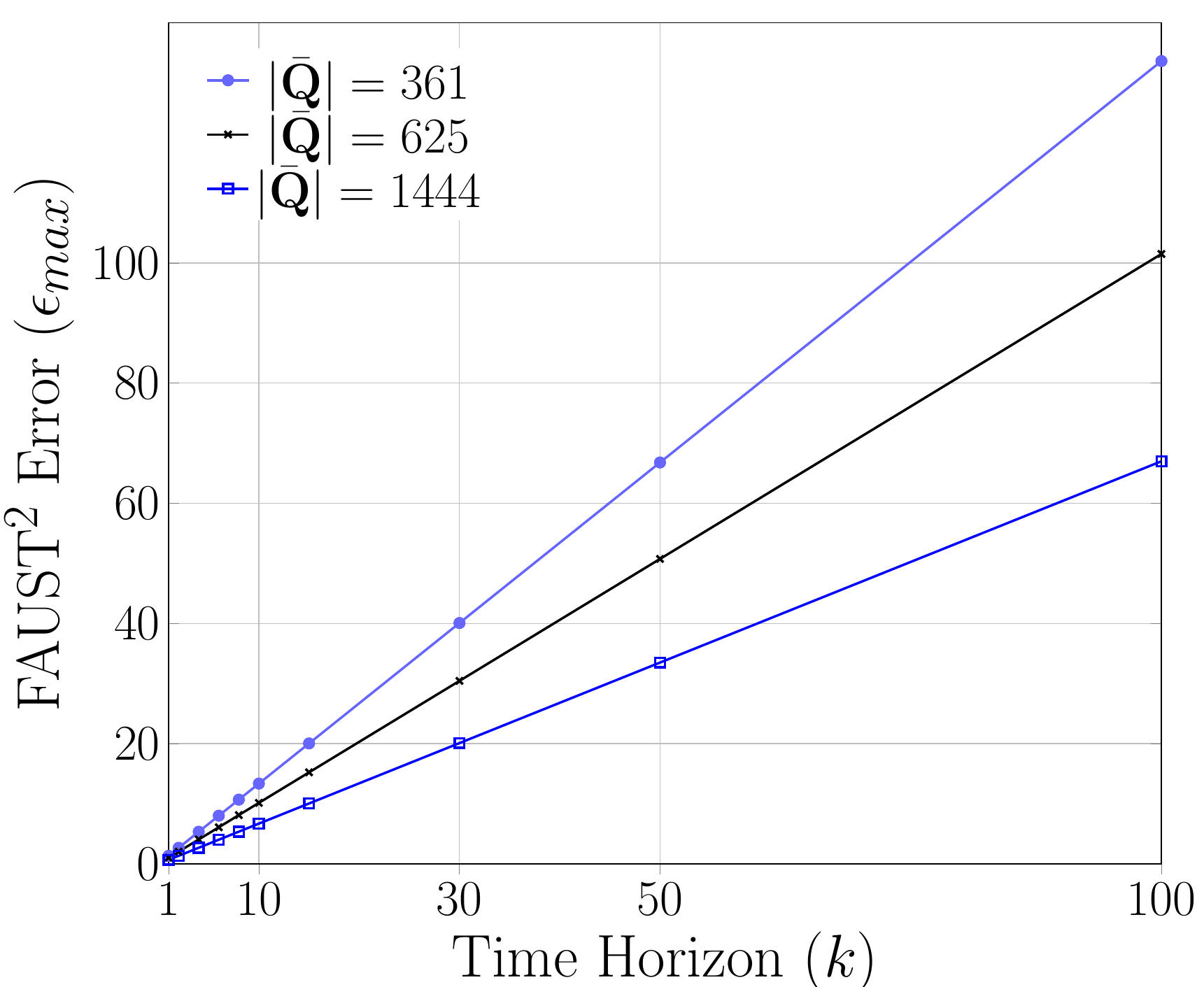}
  \caption{\faust}
  \label{fig:E_faust}
\end{subfigure}
\caption{Maximum error incurred in satisfying $\phi_1$ as a function of time horizon $k$.}
\label{fig:E_max}
\end{figure}

In Fig. \ref{fig:E_max}, we show the error of each method as a function of the time horizon $k$ in $\phi_1$.  
From these figures it is evident that our approach again greatly outperforms \faust.  That is because our method embeds the error in the abstraction and performs computations according to feasible transition probabilities, which prevents the error from exploding over time, whereas the error of \faust keeps increasing monotonically with the time horizon.  An interesting aspect in Fig. \ref{fig:E_IMDP} is that the error of our method goes to zero as $k$ increases.  That is because the system under consideration is an unbounded Gaussian process, and despite its stable dynamics, the probability of it remaining within the bounded set $\safeSet$ approaches zero as time grows larger.  This is meaningfully captured by both the upper and lower probability bounds of our method.  
On the other hand, \faust is not able to capture this behavior and its error explodes. 

\subsection{Case Study 2 - Strategy synthesis}
\label{subsec:Cs2}
\begin{figure}[t]
	\centering
\begin{subfigure}{.49\columnwidth}
    \centering
    \resizebox{3.4cm}{!}{
    \includegraphics[width=.96\columnwidth]{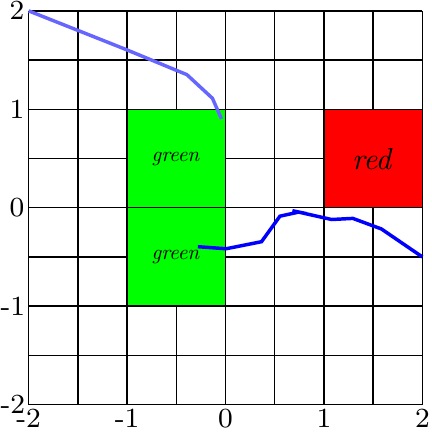}}
    \caption{}
    \label{fig:grid}
\end{subfigure}
 \begin{subfigure}{.49\columnwidth}
     \centering
    \includegraphics[width=\columnwidth]{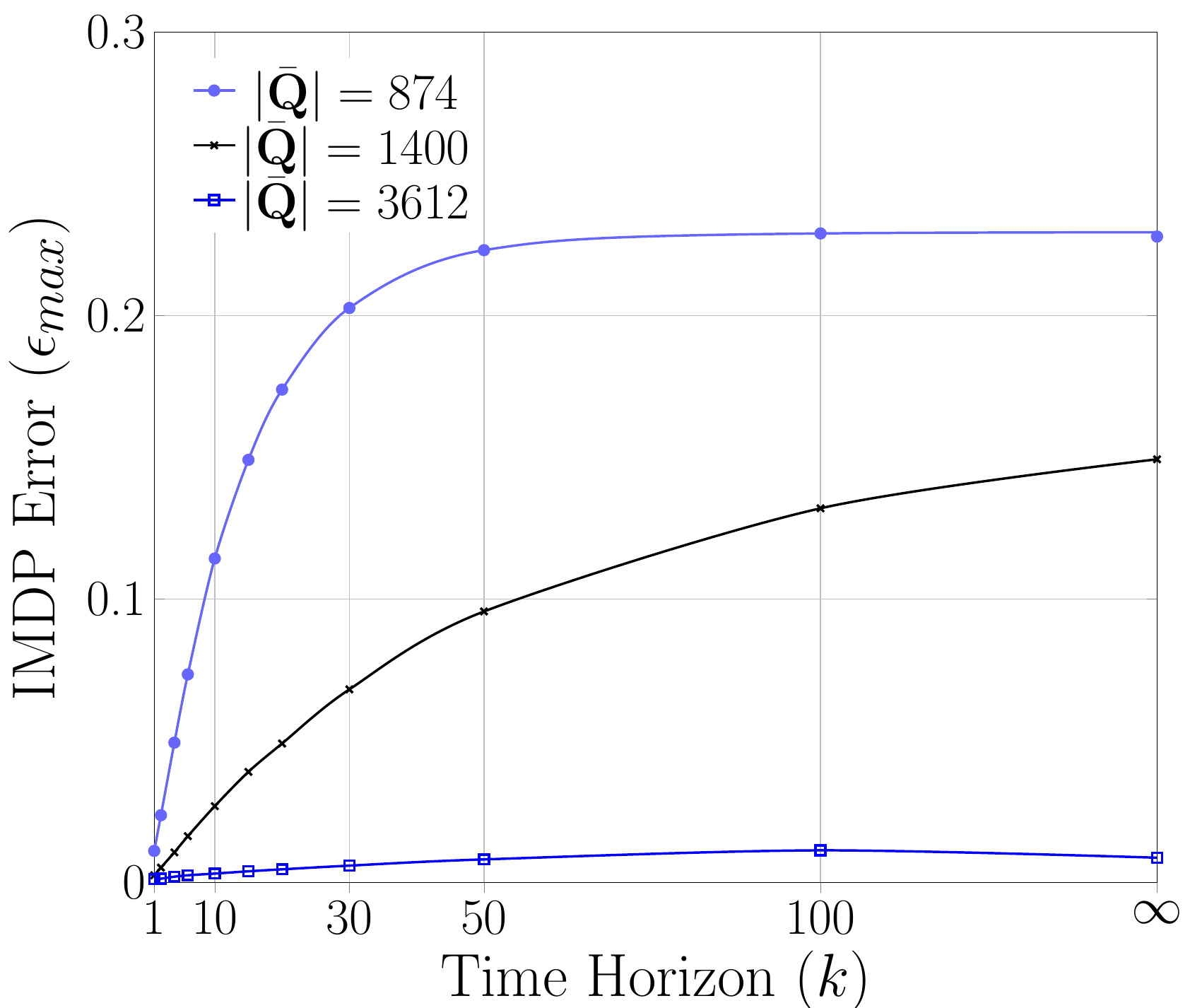}
     \caption{} 
      \label{fig:E_IMDP_SHS}
   \end{subfigure}\\ 
\begin{subfigure}{.49\columnwidth}
     \centering
     ~
      \resizebox{3.5cm}{!}{
     \includegraphics[width=.95\columnwidth]{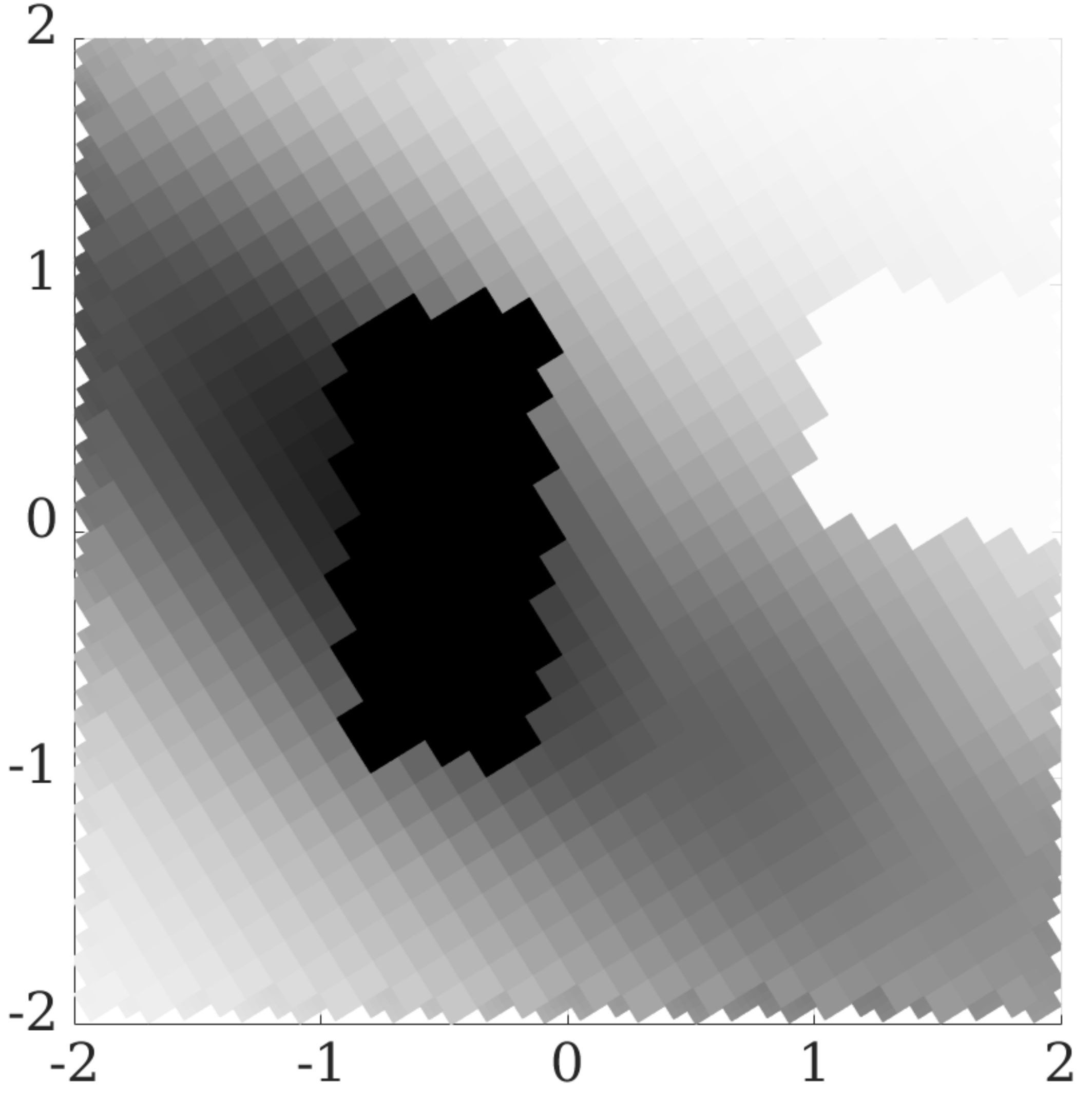}}
     \caption{} 
     \label{fig:RA_m1}
   \end{subfigure}
\begin{subfigure}{.49\columnwidth}
  \centering
   \resizebox{3.8cm}{!}{
 \includegraphics[width=1.9\columnwidth]{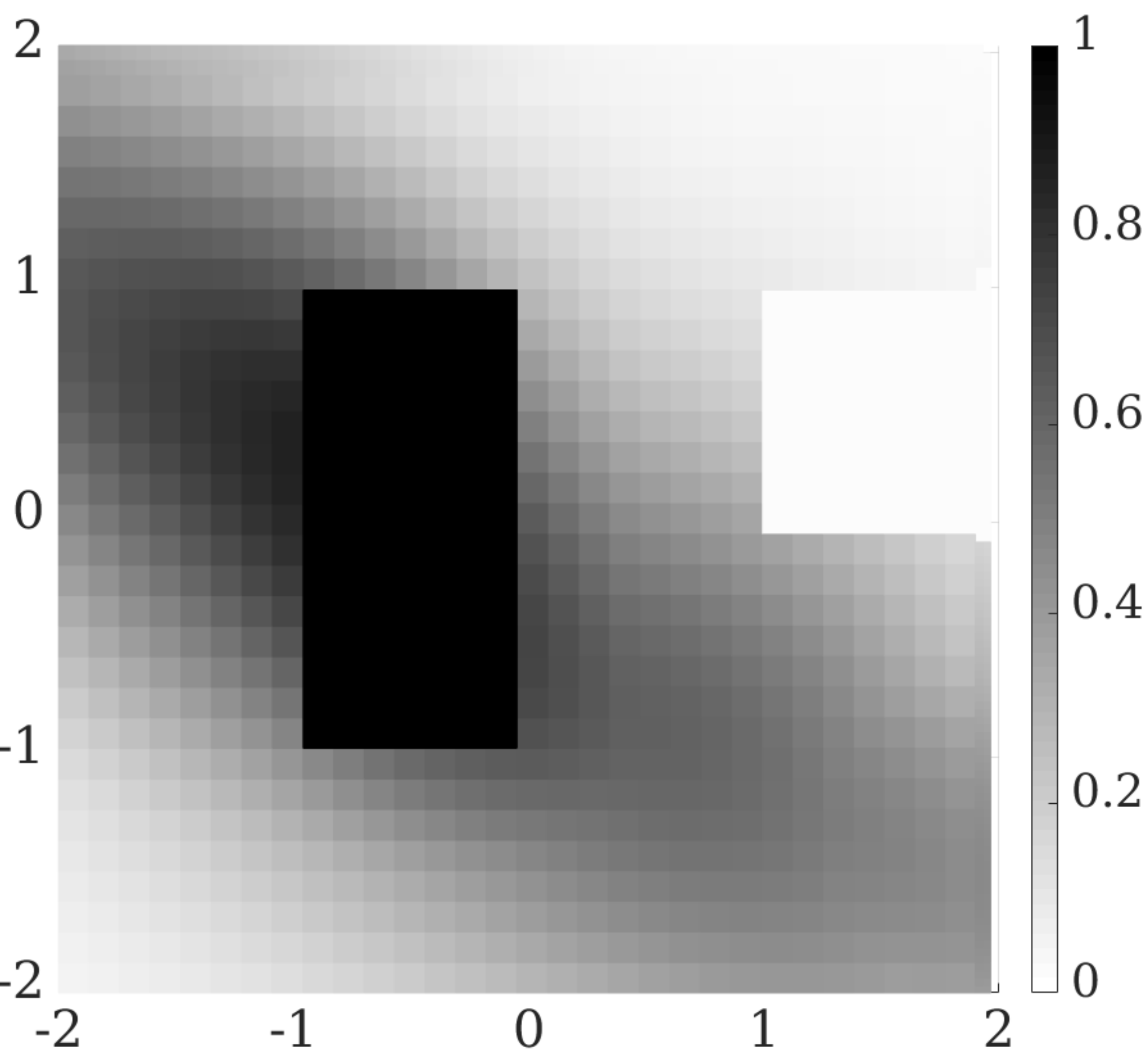}}
  \caption{ }
 \label{fig:RA_M2}
\end{subfigure}
\caption{
Synthesis results for $\phi_2$ with (a) original set $X$ with simulated trajectories under $\str^*_{\phi_2}$, (b) maximum  error  incurred  in  satisfying $\phi_2$ as function of time horizon $k$, and lower bound probabilities of satisfying $\phi_2$ for modes (c) $a_1$ and (d) $a_2$. 
}
\label{fig:Cs2_Ex2}
\end{figure}

We consider a 2-dimensional \shs with two modes $A = \{a_1, a_2\}$: 
\begin{equation*}
  F(a_1) =  
  \begin{pmatrix}
    0.1&0.9\\
    0.8& 0.2\\
 \end{pmatrix},
 \quad
 G(a_1)= 
 \begin{pmatrix}
   0.3& 0.1\\
0.1& 0.2\\
 \end{pmatrix},
\end{equation*}
\begin{equation*}
  F(a_2) =  
  \begin{pmatrix}
   0.8&0.2\\
0.1 & 0.9\\
 \end{pmatrix},
 \quad
 G(a_2)= 
 \begin{pmatrix}
    0.2 &0\\
0& 0.1\\
 \end{pmatrix}.
\end{equation*}
Note that $F(a_1)$ and $F(a_2)$ are not asymptotically stable, as they both have one eigenvalue equal to $1$. 
We are interested in synthesizing a switching strategy that maximizes the probability of satisfying
$$\phi_2= \neg {red}~ \mathcal{U} ~{green}.$$
within the set $X = [-2,2] \times [-2,2]$. The regions  associated with the labels \textit{red} and \textit{green} are depicted in Fig.~\ref{fig:grid}.

Note that $\phi_2$ has an unbounded time horizon, hence, \faust cannot be applied.
We make use of an adaptive grid, inspired by \cite{esmaeil2013adaptive}, such that the resulting cells have maximum and minimum sizes in the original space of $\dx_{\max} = 0.13$ and $\dx_{\min} = 0.05$, respectively.  
Our adaptive-grid algorithm first over-approximates $\Post(\safeSet, \transfMatrix_{a_i})$ for $i \in \set{1,2}$ by using a uniform grid with the allowed maximum-sized cells.  It refines the cells that belong to the green and red regions in the original space,  up to the resolution of the minimum-sized cells. Fig. \ref{fig:RA_m1} and \ref{fig:RA_M2} show the discretization of modes $a_1$ and $a_2$, respectively. 
The generated \imdp has $|\Qimdp| = 3612$ states with $|\Qimdp^{a_1}| = 1862$ and  $|\Qimdp^{a_2}| = 1750$. 
Note that in mode $a_1$ the cells associated with the label  $\neg {red}$ under-approximate $X \setminus {red}$, i.e., the red region is over-approximated, whereas the regions associated with the label \textit{green} under-approximate the green region.  This is due to the transformation function $\transfMatrix_{a_1}$, which includes a rotation in addition to a translation, which does not respect the regions of interest in $R$. 

We run the synthesis algorithm to obtain the robust strategy $\str^*_{\phi_2}$ with the corresponding lower probability bounds. For each state, the lower probability bounds are depicted in Fig. \ref{fig:RA_m1} and \ref{fig:RA_M2}. 
The total time to compute the abstraction and to generate $\str^*_{\phi_2}$ is 5434 seconds. 
Fig.~\ref{fig:grid} shows the simulation of two trajectories using $\str^*_{\phi_2}$ with a starting point of $(2,-0.5)$ within mode $a_1$ and $(-2,2)$ within mode $a_2$ respectively. 
In both instances, the property $\phi_2$ is satisfied.

We also analyze the errors of our method for $\phi_2$ as a function of time horizon for various grid sizes.  Fig.~\ref{fig:E_IMDP_SHS} shows the results.  It can be seen that, for a fixed $k$, $\err_{max}$ decreases monotonically with the number of states (similar to Fig. \ref{fig:E_IMDP} in Case Study 1), and $\err_{max}$ converges to a steady-state value for each grid size as the time horizon increases.  

\subsection{Case Study 3 - Scaling in continuous dimension}
\label{subsec:Cs3}
We consider a stochastic process  
with $A =\{a_1\}$ (single mode) and dynamics characterised by $F(a_1) =  -0.95\Id_{d}$ 
and $G(a_1)= 0.1\Id_{d}$, 
where $d$ corresponds to the continuous dimension of the stochastic process (number of continuous variables) and $\safeSet= [-1,1]^{d}$. 
We are interested in checking the specification 
$$\phi_3= \mathcal{G}^{\leq 50} \safeSet$$ 
as the continuous dimension $d$ of the model varies.  
We use 
a 
uniform grid characterized by parameter $\Delta x = 1$ 
per side.  
We compute the corresponding lower- and upper-bound probabilities of satisfying $\phi_3$ and 
list the number of states required for each dimension together with the associated $\err_{max}$ in Table~\ref{tab:comp_Dim}.
The method generates abstract models with manageable state spaces, and displays scalability with respect to the continuous dimension $d$ of the \shs to models with more than ten variables, which is a marked improvement over state-of-the-art tools \cite{SA15}. 

\begin{table}[h!]
    \centering
    \begin{tabular}{l|ccc}
    \textbf{Dimensions} & $|\mathbf{\bar Q}|$    & \textbf{Time taken} & \textbf{Error} \\
          (d)        & (states) & (secs)          &$(\err_{max})$   \\ \hline \hline 
     2     & 4     & 0.014 & 0.030 \\ 
     3     & 14    & 0.088 & 0.003 \\ 
     4     & 30    & 0.345 & 0.004\\ 
     5     & 62    & 1.576 &0.003\\ 
     6     & 125   & 6.150  & 0.004\\ 
     7     & 254   & 23.333 & 0.003 \\
     8     & 510   & 88.726   & 0.003\\
     9     & 1022   & 367.133 & 0.003\\
     10    & 2046   & 1787.250  &0.003 \\
     11    & 8190   &  25500.000    &0.003 \\
    \hline
    \end{tabular}
    \caption{
    Verification results of our \imdp approach for $\phi_3$.} 
    \label{tab:comp_Dim}
\end{table}

\section{Conclusions}
  \label{sec:conclusion}

This work has presented a theoretical and computational technique for analysis and synthesis of discrete-time stochastic hybrid systems. 
A suitable choice of the abstraction framework results in exact error bounds, leading to precise and compact abstractions for the synthesis tasks.  
The experimental results illustrate that the proposed framework greatly outperforms the state of the art time-wise and that is more scalable, 
thus mitigating the state-space explosion problem.  
Whilst the framework is tailored to \sstl and \ltl properties, it can be extended to verification and synthesis for more complex  and even multi-objective \cite{Hahn:QEST:2017} properties. 

\bibliographystyle{IEEEtran}
\bibliography{refs}

\newpage
\appendix

\section{Proofs}
\label{sec:proofs}
\subsection{Proof of Proposition \ref{proposition:discrete-kernel-hyperbox}}

\begin{proof}[Proof] 
       For  a fixed $a \in \U$,  recall that 
       \[\discreteKernel(q  \mid x,a)=\int_{q} \mathcal{N}( t \mid F(a) x , \cov_\pX(a) ) \, dt,\]
       where $\cov_\pX(a) = G^T(a) \cov_w G(a)$. 
    By applying a whitening through the transformation matrix $\transfMatrix_a = \Lambda_a^{-\frac{1}{2}} V_a^T$, we obtain that $\transfMatrix_a \cov_\pX(a) \transfMatrix_a^T = \Id,$
    where $\Id$ is the identity matrix. 
    Thus, by working in the transformed space induced by $\transfMatrix_a$, we obtain
    \begin{align*}
       \discreteKernel(q  \mid x,a)= 
        &\int_{\Post(q, \transfMatrix_{a})} \mathcal{N} \big(t \mid \transfMatrix_{a} \, F(a) x,\Id \big) \, dt.
    \end{align*}
    Under the assumption that $\Post(q, \transfMatrix_{a})$ is a hyper-rectangle, the above multidimensional integral can be separated and expressed as a product of $m$ integrals of uni-dimensional normal distributions:
    \begin{align*}
      \discreteKernel(q  \mid x,a) &= 
        \int_{\Post(q, \transfMatrix_{a})} \mathcal{N} \big(t \mid \transfMatrix_{a} \, F(a) x ,\Id \big) \, d t \\
        &= \int_{v_l^{(1)}}^{v_u^{(1)}} \cdots \int_{v_l^{(m)}}^{v_u^{(m)}}    \mathcal{N} \big(t_1 \mid y^{(1)}, 1 \big) \cdots \mathcal{N} \big(t_m \mid\\
        & \hspace{40mm}  y^{(m)}, 1 \big) \, d t_1 \cdots dt_m \\
        &= \prod_{i = 1}^{m} \; \int_{v_l^{(i)}}^{v_u^{(i)}} \mathcal{N} \big(t_i \mid y^{(i)}, 1 \big) \, dt_i \\
        &= \prod_{i = 1}^{m} \; \frac{1}{2} \Big( \erf(\frac{y^{(i)} -v^{(i)}_l}{\sqrt{2}}) - \erf(\frac{y^{(i)} -v^{(i)}_u}{\sqrt{2}}) \Big),
    \end{align*}
    where $y = \transfMatrix_{a} \, F(a) x$.
\end{proof}

\subsection{Proof of Theorem \ref{th:SpaceDiscretizationKKTOnditions}}
\label{sec:proof_SpaceDiscretizationKKTOnditions}

\begin{proof}[Proof]\label{ProofKKT}
We first consider the maximum case and then discuss the minimum case.
The KKT conditions guarantee that if $y\in \Post(q_i' , \transfMatrix_{a} )$ is a local maximum for $f$, then there must exist a vector of constants $\mu=(\mu_1,\ldots,\mu_k)$ such that
$\nabla f(y) ={H}^T \mu$, $\mu_i\geq 0$ for all $i\in \{1,...,k\}$, and $\mu_i(\sum_{j=1}^{m} H^{(i,j)}y^{(j)} \,-b_i )=0 $,
where $H^{(i,j)}$ is the component in the i-th row and j-th column of matrix $H$. Note that we have a constant $\mu_i$, $i \in \{1,\ldots,k \},$ for each of the half-paces defining $\Post(q_i' , \transfMatrix_{a})$.  
Thus, there are three possible cases:

    \textbf{Case 1:} \textbf{$x^*$ is not in the boundary of $\Post(q_i' , \transfMatrix_{a} )$.} In this case the KKT conditions imply that $y$ is a maximum only if  $\nabla f(y)=0.$
    For a normal distribution with identity covariance, this point is exactly $y=\big(\frac{v_{u}^{(i)}+v_{l}^{(1)}}{2},...,\frac{v_{u}^{(m)}+v_{l}^{(m)}}{2}\big).$
 If  $y \in \Post(q_i' , \transfMatrix_{a} )$, then this is the global maximum, because it is the global maximum of the unconstrained problem.

    \textbf{Case 2:}
    \textbf{$x^*$ is a vertex of $\Post(q_i' , \transfMatrix_{a} )$.} We call a vertex an intersection of $m$ half-spaces. As a consequence, we have that the KKT conditions are satisfied in $y$, vertex of $\Post(q_i' , \transfMatrix_{a} )$, if and only if  
$ \nabla f(y)=\bar{H}^T \mu ,$
    where  $\bar H$ is the submatrix that contains only the $m$ rows of $H$ representing the half-spaces interesting at $y$, and vector $\mu$ contains only the $m$ corresponding constants.
    Thus, we have a system of $m$ equations and $m$ variables that has solution for $\mu_i \in \mathbb{R}$. 
    However, since the set of vertices is finite, it is generally faster to just include all the vertices as possible candidate solutions instead of solving the system of equations.

    \textbf{Case 3:}
    \textbf{$y$ is in the boundary of $\Post(q_i' , \transfMatrix_{a} )$, but  is not a vertex.} In this case only $r<m$ of the half-spaces in $H$ intersect at $y$.
    Thus, if $y$ is a maximum then $\nabla f(y)= \bar{H}^T \mu,$
    where $\bar{H}$ is the submatrix of $H$ containing the $r<m$ half-spaces intersecting at $y$, and $\mu$ contains only the $r$ corresponding constants. 
    Note that this is a system with more equations than variables. Therefore, only when some of constraints become linearly dependent, there may be a solution for $y \in \Post(q_i' , \transfMatrix_{a} )$, if at all.
 
 The minimum case is identical except that condition $\nabla f(y) = {H}^T \mu$ is replaced with $\nabla f(y) =-{H}^T \mu$.
\end{proof}

\subsection{Proof of Proposition \ref{Proposition-LogConcaveFundaction}}
\begin{proof}
By Definition we have
\begin{align*}
    f(y) = \prod_{i=1}^m \bar{f}(y^{(i)}\mid v_l^{(i)},v_u^{(i)}),
  \end{align*}
  where 
  \begin{align*}
    \bar{f}({y^{(i)}}\mid v_{l}^{(i)},v_{u}^{(i)}) = \frac{1}{2} \big(\erf(\frac{y^{(i)}-v_{l}^{(i)}}{\sqrt{2}}) - \erf(\frac{y^{(i)}-v_{u}^{(i)}}{\sqrt{2}}) \big)
  \end{align*}
  with $v_{u}^{(i)}>v_{l}^{(i)} $.
Now,  since a product of log-concave functions is a log-concave function itself,
  to show that $f(y)$ is log-concave, it is enough to show that $\bar{f}(y^{(i)}\mid v_l^{(i)},v_u^{(i)})$ is log-concave for $i \in \{1,...,m \}$. 
 In order to do that we first need to observe that 
$$       \bar{f}(y^{(i)}\mid v_l^{(i)},v_u^{(i)})=\int_{y^{(i)}-v_u^{(i)}}^{y^{(i)}-v_l^{(i)}} \mathcal{N}(t \mid  0,1)dt. $$
That is, $\bar f$ induces a  standard Gaussian probability measure $\bar P$. We denote with $\bar P( [y^{(i)}-v_{u}^{(i)},y^{(i)}-v_l^{(i)}])$ the resulting probability for convex Borel set $[y^{(i)}-v_{u}^{(i)},y^{(i)}-v_l^{(i)}]$. By rearranging terms, for $\lambda \in [0,1], y_1,y_2 \in \mathbb{R}$, we finally obtain
       \begin{align*}
       &\bar f(\lambda y_1 + (1-\lambda) y_2 \mid v_l^{(i)},v_u^{(i)})=\\
      & \bar P(\lambda [ y_1 -v_{u}^{(i)}, y_1- v_l^{(i)}]+  (1-\lambda)[ y_2-v_{u}^{(i)}, y_2-v_l^{(i)}])\geq\\
       & \bar P([ y_1-v_u^{(i)},y_1 -v_l^{(i)}])^{\lambda} \bar P[ y_2-v_u^{(i)},y_2-v_l^{(i)}])^{1-\lambda}=\\
       &\bar f(y_1 \mid v_l^{(i)},v_u^{(i)})^{\lambda} \bar  f(y_2  \mid v_l^{(i)},v_u^{(i)}))^{(1-\lambda)}, 
       \end{align*}
       where the above inequality is due to Theorem $2$ in \cite{prekopa1971logarithmic}.
\end{proof}

\subsection{Proof of Proposition \ref{prop:tran_unsafe}}
\label{sec:proof_tran_unsafe}

\begin{proof}[Proof]
    For the upper bound, we have that for $q_i\in \Qimdp_\safe$ and $a \in \Amdp$,
    \begin{align*}
       \max_{x \in \qimdp_i}  \, \discreteKernel(X \mid x, \aimdp) 
       & \leq \max_{x \in q_i} \int_{\safeSet} \mathcal{N}(z \mid  F(a)x, \cov_{\pX}(a) \,dz \\
       & = \max_{y\in \Post(q_i', \transfMatrix_a)} \int_{\Post(\safeSet, \transfMatrix_a)}\mathcal{N}(z \mid y,\Id )\,dz \\
       & \leq \max_{y\in \Post(q_i', \transfMatrix_a)} \sum_{q\in \bar \Qimdp^a}\int_{\Post(q, \transfMatrix_a)} \N (z \mid y, \Id ) \,dz\\ 
       & = \max_{y \in \Post(q_i', \transfMatrix_a)} \sum_{q \in \bar \Qimdp^a} f(y,q).
    \end{align*}


    For the lower bound, similarly to the upper bound, we have that
    \begin{equation*}
        \min_{x \in q_i} \discreteKernel(\safeSet \mid x,a)\geq 
         \min_{y \in \Post(q_i',  \transfMatrix_a)}\sum_{q \in  \Qimdp^a} 
        f(y,q).
    \end{equation*}
    
\end{proof}

\subsection{Proof of Theorem \ref{th:Correctness}}

    For each $\phi$, let $\Adfa_{\bar\phi}=(\Zdfa, 2^{\bar\AP}, \Trandfa, \zdfa_0, \Fdfa)$ be the \dfa correspondent to $\phi$ with initial state $z_0.$
    Then, $P(\phi   \mid x, X, \strH^*)$ can be computed on the product stochastic hybrid system $\H_{\phi}= \H \times \Adfa_{\bar\phi} =(A\times \Zdfa, F_{\phi}, G_{\phi}, \AP, L_{\phi})$, where  $L_{\phi}(x,(a,z))=L((a,x)),\, F_{\phi}(a,z)=F(a)$ and $G_{\phi}(a,z)=G(a)$. We define the set of accepting states of $\H_\phi$  as $X_{ac}=X\times A \times \Fdfa$.
    It is possible to show that  $P(\phi   \mid x, X, \strH^*)$ 
    can be computed as the solution of the following Bellman equation
         \begin{align}
         \nonumber
        V(&z_0,x, X, \strH^*)=\\
        &\begin{cases}
        1 \quad \text{if } (x,\strH^*(x),z_0) \in X_{ac}\\
        0 \quad \text{if }x \not\in X\\
        \int_{X}  f(x' |  x, \strH^*(x_0))  V( \Trandfa(z_0,L(x,\strH^*(x)),x', X, \strH^*) dx\end{cases}
\label{eq:BellmanSHS}
    \end{align}
  where  $f(x'| x,$ $\strH^*(x_0)) $ the density function of transition kernel $T$ and, with an abuse of notation, we call $\strH^*(x_0)$ the action resulting from the application of the (stationary) strategy $\strH^*$ in $x_0.$
 For $q\in Q$ call $$\breve V^{\strH^*}(z,q, X, \strH^*)=\min_{x \in q} V( z,x, X, \strH^*).$$
Then, it follows that 
\begin{align*}
   &\breve V^{\strH^*}(z_0,q, X, \strH^*) =\\
   &\begin{cases}
        1 \quad \text{if there exists $x\in q$ s.t. }(x,\strH^*(x),z_0) \in X_{ac}\\
         0 \quad \text{if }x \not\in X\\
      \min_{x \in q}  \int_{X}  f(x' |  x, \strH^*(x))  V( \Trandfa(z_0,L(x,\strH^*(x)),x', X, \strH^*) dx'\end{cases}
\end{align*}
Then, because for each $x_1,x_2 \in q$ it holds that $\strH^*(x_1)=\strH^*(x_2)$ and  $Q_{\phi}$ is a discretization of $X$ that respects the propositional regions, we obtain
\begin{align*}
   &\breve V^{\strH^*}(z_0,q, X, \strH^*) \leq \\
   &\begin{cases}
        1 \quad \text{if there exists $x\in q$ s.t. }(x,\strH^*(x),z_0) \in X_{ac}\\
         0 \quad \text{if }x \not\in X\\
    \min_{x \in q}   \sum_{q \in Q_{\phi}}  T(q |  x, \strH^*(x)) 
    \breve V^{\strH^*}(\Trandfa(z_0,L(x,\strH^*(x)),x', X, \strH^*)\end{cases}
\end{align*}
   The latter expression is exactly \eqref{eq:bellman_lower} for a fixed strategy $\strH^*.$ 
   Similar approach can be used to prove that the solution of \eqref{eq:BellmanSHS} is upper bounded by  \eqref{eq:bellman_upper}.

\subsection{Proof of Lemma \ref{prop:upper_bound}}
$Q_{\phi}$ is a discretization of $X$ that does not respect the propositional regions $R$, and the labeling function $L$ of $\I$ introduces an under approximation of those regions. 
Similar to the proof of Theorem \ref{th:Correctness}, a product \shs $\H_\phi$ can be constructed.  By replacing the discretization $Q_\phi$ in the Bellman equation and noting that $L$ under-approximates $R$, it holds that $\breve V^{\strH^*}(z_0,q, X, \strH^*)$ is an under-approximation of $P(\phi \mid s_0, X, \strH^*)$.

For the upper bound, note that the labeling function $L'$ over-approximates the labels of each region.  With the same derivation as above but using $L'$ instead of $L$, it follows that 
$$\hat V^{\strH^*}(z,q, X, \strH^*) \geq P(\phi \mid s_0, X, \strH^*),$$
where
$$\hat V^{\strH^*}(z,q, X, \strH^*)=\max_{x \in q} V( z,x, X, \strH^*),$$ 
and 
$V( z,x, X, \strH^*)$ is defined in \eqref{eq:BellmanSHS}.

\section{Case Study 1}
\label{app:Cs1}
\begin{figure}[h!]
	\centering
\begin{subfigure}{.49\columnwidth}
     \centering
     \includegraphics[width=.86\columnwidth]{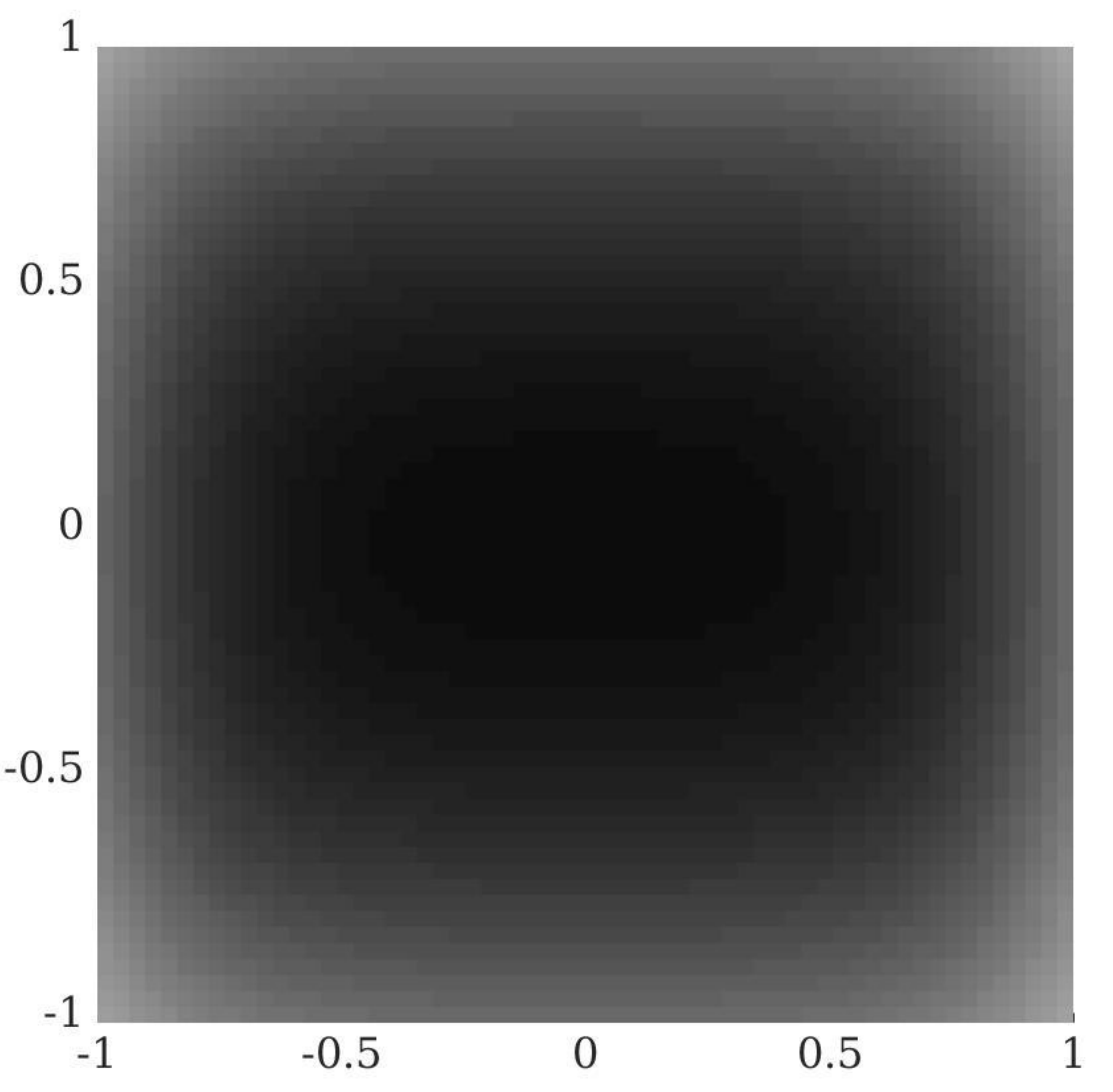}
     \caption{\imdp} 
     \label{fig:imdp_lower}
   \end{subfigure}
\begin{subfigure}{.49\columnwidth}
  \centering
 \includegraphics[width=.95\columnwidth]{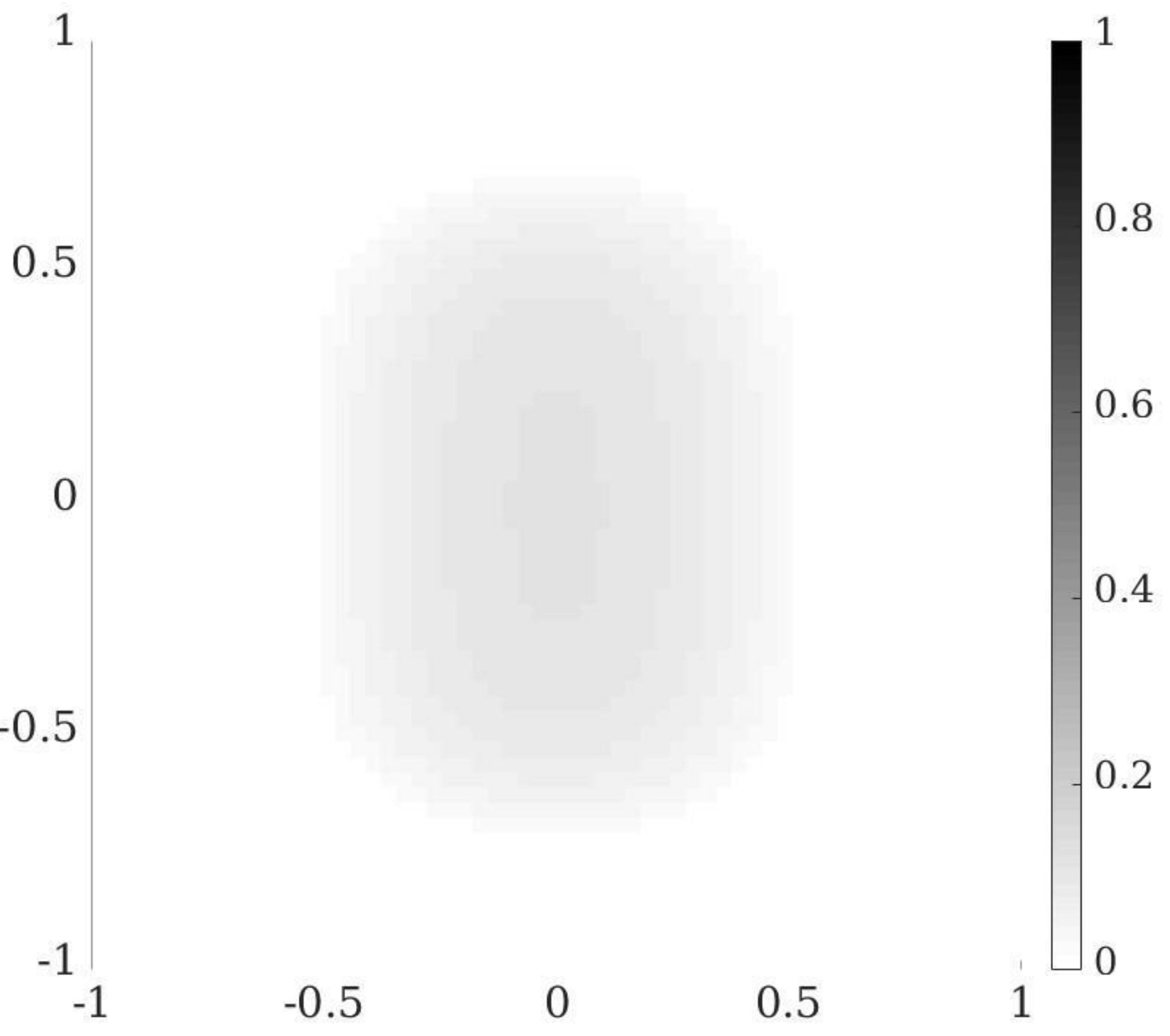}
  \caption{\faust}
  \label{fig:faust}
\end{subfigure}
\caption{Lower bound probabilities of satisfying $\phi_1$ with $|\bar \Qimdp|$ = 3721 and $k = 2$.}
\label{fig:Cs1_Ex2}
\end{figure}

We present the lower bound probabilities of satisfying $\phi_1$ using both \imdp and \faust based abstractions, for the particular grid $|Q|=3722$ in Fig.~\ref{fig:Cs1_Ex2}.
This further highlights that our approach greatly outperforms the state of the art with respect to probability of satisfaction 
for the same size of the grid. 

\end{document}